\documentclass[11pt,fleqn]{article}
\usepackage{a4wide}
\usepackage{graphicx}
\usepackage{longtable}
\usepackage{times}
\usepackage{epsfig}
\usepackage{amsmath}
\usepackage{amssymb}
\usepackage{cite}

\makeatletter
\renewcommand{\fnum@table}{\textbf{\tablename~\thetable}}
\renewcommand{\fnum@figure}{\textbf{\figurename~\thefigure}}
\makeatother

\newcounter{myenumi}

\renewcommand{\themyenumi}{\roman{myenumi}}
{\end{list}}

\newlength{\myem}
\newcounter{mysubequation}[equation]

\newcommand{\cpv}{CP violation}
\makeatletter
\renewcommand{\section}{\@startsection{section}{1}{0em}{-\baselineskip}%
{\baselineskip}{\normalfont\large\bfseries}}
\renewcommand{\subsection}%
{\@startsection{subsection}{2}{0em}{-0.7\baselineskip}%
{0.7\baselineskip}{\normalfont\bfseries}}
\makeatother

\newcommand{\ie}{{\it i.e.}}

\newcommand{\eg}{{\it e.g.}}

\newcommand{\fig}{Figure}

\newcommand{\abr}[1]{{\sc\lowercase{#1}}}

\newcommand{\dm}[1]{\Delta m_{#1}^2}

\newcommand{\bi}{\begin{itemize}}
\newcommand{\ei}{\end{itemize}}

\newcommand{\ver}[1]{#1}

\newcommand{\theorefs}{Mohapatra:2002kn,Chen:2003zv,Mohapatra:2003qw,Pati:2003qi,Altarelli:2003vk,APStheo}

\newcommand{\bea}{\begin{eqnarray}}
\newcommand{\eea}{\end{eqnarray}}
\newcommand{\be}{\begin{equation}}
\newcommand{\ee}{\end{equation}}

\newcommand{\numu}{\mbox{$\nu_{\mu}$}}
\newcommand{\nutau}{\mbox{$\nu_{\tau}$}}

\newcommand{\nue}{\mbox{$\nu_{e}$}}

\def\nue{\ensuremath{\nu_{e}}}
\def\nubare{\ensuremath{\overline{\nu}_{e}}}

\def\numu{\ensuremath{\nu_{\mu}\ }}
\def\nubarmu{\ensuremath{\overline{\nu}_{\mu}}}
\def\nubartau{\ensuremath{\overline{\nu}_{\tau}}}

\def\nutau{\ensuremath{\nu_{\tau}\ }}

\newcommand{\nuenumu}{\ensuremath{\nue \rightarrow \numu\,}}
\newcommand{\numunutau}{\ensuremath{\numu \rightarrow \nutau\,}}
\newcommand{\nuenutau}{\ensuremath{\nue \rightarrow \nutau}}

\newcommand{\nubarmunubare}{\ensuremath{\overline{\nu}_\mu \rightarrow \overline{\nu}_e\,}}

\newcommand{\dmtt}{\ensuremath{\Delta m^2_{23} \,}}

\newcommand{\He}{\ensuremath{^6{\mathrm{He}\,}}}
\newcommand{\Ne}{\ensuremath{^{18}{\mathrm{Ne}\,}}}

\newcommand{\thetaot}{\ensuremath{\theta_{13}}\,}
\newcommand{\thetatt}{\ensuremath{\theta_{23}}\,}
\newcommand{\numunue}{\ensuremath{\nu_\mu \rightarrow \nu_e}}

\newcommand{\sigdm}{\ensuremath{{\rm sign}(\Delta m^2_{23})\ }}
\newcommand{\delCP}{\ensuremath{\delta_{\rm CP}\ }}

\newcommand{\stheta}{\sin^22\theta_{13}}
\newcommand{\deltacp}{\delta_\mathrm{CP}}

\newcommand{\capdef}{}
\newcommand{\mycaption}[2][\capdef]{\renewcommand{\capdef}{#2}%
       \caption[#1]{{\it #2}}}
\begin{document}
\def\baselinestretch{1.05}

\begin{center}
{\Large \bf Future neutrino oscillation facilities} \\
\vskip12pt \noindent
 {\large A.~Blondel$^a$, A. Cervera-Villanueva$^a$, A. Donini$^b$,  P.~Huber$^c$, 
 M.~Mezzetto$^d$, P.~Strolin$^e$ }\\
\end{center}

\noindent
{\it
 {$^a$  Section de Physique, Universit\'e de Gen\`eve, Switzerland} \\
 {$^b$  I.F.T. and Dep. F\'isica Te\'orica, U.A.M., Madrid, Spain} \\
{$^c$  Department of Physics, University of Wisconsin } \\
 {$^d$ INFN, Sezione di Padova, Italy} \\
 {$^e$ Universit\'a degli Studi e Sezione INFN, Napoli, Italy}
} 

\vskip24pt
\begin{abstract}
 The recent discovery that neutrinos have masses opens a wide 
new field of experimentation. Accelerator-made neutrinos are essential
 in this program. Ideas for future facilities include high intensity
 muon neutrino beams from pion decay (`SuperBeam'), electron neutrino
 beams from nuclei decays (`Beta Beam'), or muon and electron neutrino
 beams from muon decay (`Neutrino Factory'), each associated with one
 or several options for detector systems. Each option offers synergetic
 possibilities, e.g. some of the detectors can be used for proton decay
 searches, while the Neutrino Factory is a first step towards muon colliders.
 A summary of the perceived virtues and shortcomings of the various options,
 and a number of open questions are presented.
\end{abstract}

\section{ Physics of massive neutrinos}

\subsection{Status}
Neutrino physics has become one of the most active areas of research
in particle physics. There are many reasons for this development, one
of the most important is that neutrino physics is a data driven field
-- for several years now, new data are pouring at an astounding rate.
It began in \oldstylenums{1987} with the first detection of supernova
neutrinos.  Although only $19$ events were
observed~\cite{Hirata:1987hu,Bionta:1987qt}, they allowed to confirm
the standard picture of core-collapse supernov\ae. Furthermore those
$19$ events also constitute the detection of the oldest neutrinos
ever; they were produced some $150\,000$ years ago. The fact that
there were detectable neutrinos after this time allows to put
stringent bounds on the neutrino life time. Furthermore the
environment at the production site was a very special one -- a dense
and hot proto-neutron star. This offers the possibility to derive
strong bounds on any additional interaction neutrinos could have.  A
comprehensive review on the properties of neutrinos which can be
deduced from supernov{\ae} is given in~\cite{Raffelt:1999tx}.

In \oldstylenums{1998} \ver{Super-K's} atmospheric neutrino
data~\cite{Fukuda:1998mi} gave the first clear evidence for neutrino
oscillation. This result was a real turning point for neutrino
physics.  Neutrino oscillation implies that neutrinos do have a mass
and the finding that the mixing angle is large was completely
unexpected.  In analogy to the quark sector the common belief was that
if neutrinos mixed at all then the mixing angles should be small. The
importance of the \ver{Super-K} result is that it is the first strong
evidence for physics beyond the Standard Model. With the \ver{Super-K}
result the number of publications per year containing the word
neutrino in their title four-folded\footnote{According to the {\tt
    hep-ph} preprint server of {\tt arXive.org}}.

The year \oldstylenums{2002} was an annus mirabilis for neutrino
physics.  The solar neutrino puzzle was proven to be due to the
properties of the neutrino and not of the Sun. The neutral current
data of \ver{SNO}~\cite{Ahmad:2002jz} yielded an independent
determination of the total flux of active neutrinos from the Sun and
in combination with other solar neutrino data proved that solar
neutrinos undergo a flavor transition. \ver{Kamland}~\cite{Kamland}
provided an independent check of the oscillation hypothesis by using
reactor neutrinos and constrained the mixing parameters to the so
called \abr{lma} solution.  These two results together are extremely
difficult to explain other than by neutrino oscillation. Also the
evidence for oscillation in atmospheric neutrinos has been confirmed
independently by \ver{K2K}~\cite{Ahn:2002up,Aliu:2004sq}, which is the
first long baseline experiment.  Furthermore two pioneers of neutrino
physics were awarded the Nobel prize.  Masatoshi Koshiba was awarded
one fourth of the prize for the detection of neutrinos from a
supernova and Ray Davis Jr. was awarded another fourth for his
detection of solar neutrinos.  A relatively recent review on the topic
of neutrino oscillations in general is given in
\eg~\cite{Gonzalez-Garcia:2002dz}.

 The above experiments indicate the presence of two mass splittings
$\Delta m^2_{21}$ and $\Delta m^2_{31}$, corresponding to solar
and atmospheric neutrino oscillations, in agreement with existence of three
active neutrinos.
There has been a further experiment
observing evidence for neutrino oscillation -- \ver{LSND}. The results
of this experiment indicate that there is a third mass splitting
$\dm{\mathrm{LSND}}$ in the range $0.2 -
10\,\mathrm{eV}^2$~\cite{Aguilar:2001ty}. The \ver{Karmen}
experiment~\cite{Armbruster:2002mp} on the other hand excludes a large
part of the parameter region claimed by \ver{LSND}. In a combined
analysis of both data sets there still remains a combined allowed
region~\cite{Wolf:2001gu}. The third mass splitting cannot be
accommodated within a three neutrino flavor framework. Basically two
possible solutions exist -- either there are more than three
neutrinos, which means that the additional neutrinos are sterile in
order not to create a conflict with the decay width of the $Z^0$ (see
\eg~\cite{PDG}), or there is a huge \abr{cpt}-violation, which would
make the mass splittings of neutrinos and of anti-neutrinos
independent of each other.  Both solutions suffer from
phenomenological problems, \ie\ they do not fit the existing data very
well. Finally \ver{Miniboone} will thoroughly test the results of
\ver{LSND}.  A fairly recent review of neutrino physics is to be found
in~\cite{Barger:2003qi}, which also contains an extensive
bibliography.

The current parameters are roughly
\begin{eqnarray}
\dm{21}\sim 8\cdot 10^{-5}\,\mathrm{eV}^2 \,\,&\mathrm{and}&\,\, \theta_{12}\sim1/2\nonumber\\
\dm{31}\sim 2\cdot 10^{-3}\,\mathrm{eV}^2\,\,&\mathrm{and}&\,\,\theta_{23}\sim\pi/4\nonumber\\
&\theta_{13}&\lesssim0.15
\end{eqnarray}
This implies a lower bound on the mass of the heaviest neutrino
$\sqrt{2\cdot 10^{-3}\,\mathrm{eV}^2} \sim 0.04 \,\mathrm{eV}$ but we
currently do not know which neutrino is the heaviest. For a more
detailed global analysis of the data see \eg~\cite{Fogli:2005gs}.  We
now can contrast our knowledge of the neutrino mixing with the quark
sector
\begin{equation}
U_{CKM}=\left(\begin{array}{ccc}
1&0.2&0.005\\
0.2&1&0.04\\
0.005&0.04&1
\end{array}\right)
\quad\quad
U_\nu=\left(\begin{array}{ccc}
0.8&0.5&?\\
0.4&0.6&0.7\\
0.4&0.6&0.7
\end{array}\right)
\end{equation}
The mixing of neutrinos is very different from that of quarks, since
there are two large mixing angles.  Neutrino masses are also peculiar
because they are at least five orders of magnitude smaller than the
mass of the electron.  These facts pose a major challenge to any
theory of neutrino masses and mixings: Why are neutrino masses so
small? Why is the neutrino mixing pattern so different from that of
the quarks? What is the pattern of neutrino masses?

\subsection{Origin of neutrino mass}

The Standard Model
is in a paradox situation -- it is extremely successful in describing 
elementary particles and their interactions and still it is strongly believed
to be incomplete. It seems to be the correct 
description of the physics which can be observed
at low energies but it is obvious from its structure
that it cannot be correct up to the very highest energies. 
This has inspired many attempts to provide a convincing model for the physics
beyond the Standard Model. These attempts strongly suffered from the fact that
no deviation from the Standard Model had been found before the discovery
of neutrino flavor transitions.  The discovery of 
neutrino oscillations is the first unequivocal experimental result
which is beyond the SM. Neutrinos within the SM are strictly massless,
\ie\ it is impossible to write down a mass term for the neutrino which
is gauge invariant and renormalizable. 

One possibility to approach this problem can be formulated in the
language of effective field theory. The SM is believed to be an
effective field theory, which means it is only valid up to some
energy scale $\Lambda$. At this scale new physics and degrees of
freedom will appear. In the absence of a full theory for the new
physics one still can write down a parameterization of the effects of
the new physics in terms of non-renormalizable (within the SM)
operators
\begin{equation}
\mathcal{L}_{SM}+\frac{1}{\Lambda}\mathcal{L}_5+\frac{1}{\Lambda^2}\mathcal{L}_6+\ldots
\end{equation}
where the higher dimensional operators are suppressed by increasing
powers of $\Lambda$. In general there can be a large number of
operators of a given dimension, \eg\ there are many dimension 6
operators, among them is the one responsible for proton decay. In that
picture the first correction to the SM is expected to come from the
dimension 5 operators $\mathcal{L}_5$ since they are only suppressed by
one power of $\Lambda$. It turns out that there is only one dimension 5
operator, given by
\begin{equation}
\mathcal{L}_5=\frac{1}{\Lambda}(LH)(LH)\rightarrow\frac{1}{\Lambda}(L\langle
H\rangle)(L\langle H \rangle)\,.
\end{equation}
This operator is the neutrino mass operator. The fact that neutrino
mass is the first correction to the SM which is expected from theory
and found in experiment is a interesting coincidence.

Technically it is straightforward to generate the neutrino mass
operator by introducing new degrees of freedom. The most commonly used
choice are heavy right handed neutrinos $N_R$, which are singlets under
the SM gauge group. In that case it becomes possible to write down a
Majorana mass for the neutrinos
\begin{equation}
\mathcal{L_\nu}=m_D\overline{\nu_L}N_R+\frac{1}{2}m_R \overline{N^c_L} N_R + h.c.\,.
\end{equation}
This construction yields light neutrino masses of the order
\begin{equation}
m_\nu\simeq \frac{m^2_D}{m_R}\,,
\end{equation}
which is the famous seesaw relation. Putting a typical fermion mass of
$m_D=100\,\mathrm{GeV}$ and $m_R=10^{15}\,\mathrm{GeV}$ at the GUT scale
yields a neutrino mass of order $0.01\,\mathrm{eV}$. This value is
tantalizing close to the order of magnitude indicated by oscillations.
In this scheme the smallness of neutrino masses is natural consequence
of the heaviness of the right handed neutrino.  In such a scenario
neutrino masses are a probe of very high energy scales which may
otherwise be not accessible. It turns out that it is far from trivial
to construct a theory which can account for the observed mixing
pattern, \ie\ predict two large mixing angles. The seesaw mechanism is
just one example and there a plethora of other possibilities. For some reviews on the
vast amount of literature on these topics see~\cite{\theorefs}.

\subsection{Baryogenesis}

The observable Universe only contains matter and no anti-matter. This
is a very surprising experimental fact since the initial condition are
thought to be symmetric with respect to matter and anti-matter.
Baryogenesis aims at finding an explanation for the observed matter
anti-matter asymmetry. It turns out that within the SM model it is in
principle very well possible to create some asymmetry but the numerical
value is way too small. For that reason this asymmetry points to
physics beyond the SM. More surprisingly, the same new physics which
is invoked to explain neutrino masses may be at the heart of baryogenesis.
Assuming that the Universe was hot enough a some point in its history
to keep $N_R$ in thermal equilibrium there will be a vast abundance of
$N_R$. The $N_R$ will fall out of equilibrium during the evolution and
finally decay. This decay can be CP-violating and therefore produce a
net lepton number
\begin{equation}
\Gamma(N_R\rightarrow L H)-\Gamma(N_R\rightarrow \overline{L} H^*)\neq
0.
\end{equation}

This lepton number later on will be converted to baryon number by
non-perturbative processes. 
\subsection{Phenomenological consequences}

In the context of GUT scale right handed neutrinos it is very
difficult to establish a one-to-one correspondence between high and
low-energy observables. A given model, however, usually has generic
predictions for low energy observables. Therefore studying neutrinos
allows to gain considerable insight into phenomena which otherwise
would be inaccessible. Colliders can not probe this kind of physics,
since any effects in scattering amplitudes are suppressed by
$m_{GUT}$, at LHC this would be effects of $\mathcal{O}(10^{-10})$!
In general any model of neutrino mass should provide predictions for
the mixing pattern and the mass scale as well as whether neutrinos are Majorana
or Dirac particles. In  terms of flavor physics and the quest for a
theory of flavor neutrinos provide one half of the available data.

\begin{table}
\mycaption{\label{tab:textures} Predictions for certain oscillation related quantities from
  various textures of the neutrino mass matrix under the assumption of
  a diagonal lepton mass matrix. Adapted from~\cite{APStheo}.}
\begin{tabular}{|c|c|c|c|c|c|} \hline
Case & Texture & $\,$Hierarchy$\,$ & $|U_{e3}|$ &
$\,|\cos2\theta_{23}|$ & $\,$Solar Angle$\,$ \\ \hline
A & $\frac{\sqrt{\Delta m^2_{13}}}{2} \left(\begin{array}{ccc}0&0&0\\
0&1&1 \\ 0&1&1\end{array}\right)$& Normal &$\sqrt{\frac{\Delta
m^2_{12}}{\Delta m^2_{13}}}$ &  $\sqrt{\frac{\Delta
m^2_{12}}{\Delta m^2_{13}}}$ & ${\cal{O}}(1)$ \\ \hline
B &  $\sqrt{\Delta m^2_{13}}\left(\begin{array}{ccc}1&0&0\\
0&\frac{1}{2}&-\frac{1}{2} \\
0&-\frac{1}{2}&\frac{1}{2}\end{array}\right)$& Inverted & $\frac{\Delta
m^2_{12}}{|\Delta m^2_{13}|}$ & $\frac{\Delta m^2_{12}}{|\Delta
m^2_{13}|}$ & ${\cal{O}}(1)$ \\ \hline
C &  $\frac{\sqrt{\Delta m^2_{13}}}{\sqrt{2}}
\left(\begin{array}{ccc}0&1&1\\ 1&0&0\\ 1&0&0\end{array}\right)$ &
Inverted & $\frac{\Delta m^2_{12}}{|\Delta m^2_{13}|}$ &
$\frac{\Delta m^2_{12}}{|\Delta m^2_{13}|}$ &
$\begin{array}{c}
 |\cos2\theta_{12}| \\ \sim\frac{\Delta m^2_{12}}{|\Delta m^2_{13}|}
\end{array}$
\\ \hline
Anarchy & $\sqrt{\Delta m^2_{13}}\left(\begin{array}{ccc}1&1&1\\ 1&1&1 \\
1&1&1\end{array}\right)$ & Normal
& $>0.1$ & -- &  ${\cal{O}}(1)$ \\ \hline
\end{tabular}
\end{table}

Based on consideration like the one summarized in
table~\ref{tab:textures} it is possible to identify a set of key
measurements necessary to identify credible scenarios for neutrino
mass generation. Another example are $SO(10)$ GUT models, which
typically make quite precise predictions for $|U_{e3}|$. The most
sensitive low energy observables and the environment to measure them
are

\begin{itemize}
\item Majorana vs Dirac mass -- $0\nu\beta\beta$
\item Absolute $m_\nu$ -- Katrin, Cosmology
\item How large is $\theta_{13}$? -- Oscillation
\item Which one is the heaviest neutrino? --$0\nu\beta\beta$, Katrin, Oscillation
\item Is $\theta_{23}$ maximal? -- Oscillation
\item Is there leptonic CP violation? -- Oscillation
\item Are there only 3 light neutrinos? -- Oscillation
\end{itemize}

If MiniBooNE should find evidence for a forth neutrino the last item
would move up to number one and it would change our view on neutrinos
and model building profoundly. What is remarkable about that list is
that a large number of items can be studied by neutrino oscillation.

Massive neutrinos offer a variety of fascinating new phenomenology
beyond oscillation. A rather
extensive review on the wide area of neutrinos in cosmology is given
in \cite{Dolgov:2002wy}. Cosmology has undergone a tremendous increase
in the available experimental data as well and is now entering a phase
of high precision measurements. With the data of
\ver{WMAP}~\cite{Wright:2003qm} and the
\ver{2dFGRS}~\cite{Percival:2002gq} the cosmological limits on the
masses of neutrinos are already slightly better than the laboratory
bounds, see \eg~\cite{Hannestad:2003xv}.

In order to develop and finally test a theory of neutrino masses and
mixing it will be essential to further improve on the knowledge of
not only the oscillation parameters but also the absolute mass scale
of neutrinos. Furthermore the observation of neutrino-less double
$\beta$-decay could shed some light onto the Majorana nature of
neutrinos. The theoretical motivation, the current status and future
experiments for neutrino-less double $\beta$-decay are reviewed
in~\cite{Vogel:2000vc,Elliott:2002xe}, whereas the prospects of
determining the absolute mass scale are reviewed
in~\cite{Bilenky:2002aw}.

\subsection{Neutrino oscillation}

The mass eigenstates are related to flavor eigenstates by $U_\nu$,
thus a neutrino which is produced as flavor eigenstate is a
superposition of mass eigenstates. These mass eigenstates propagate
with different velocity and a phase difference is generated. This
phase difference gives rise to a finite transition probability
\begin{equation}
P_{\nu_\alpha\rightarrow\nu_\beta}=\sum_{ij} U_{\alpha j} U^*_{\beta
  j} U^*_{\alpha i} U_{\beta i} e^{-i\frac{\Delta m^2_{ij}L}{2E}}
\end{equation}
Neutrino oscillation is a quantum mechanical interference phenomenon
and therefore it is uniquely sensitive to extremely tiny effects.

In order to get some qualitative understanding it is useful to use the
two flavor approximation, \ie\ only one mass splitting $\dm{}$ and one
mixing angle $\theta$. In that case the oscillation probabilities take
the simple form
\begin{eqnarray}
P_{\nu_\alpha\rightarrow\nu_\beta}&=&\sin^2 2 
\theta\sin^2\left(\frac{\dm{}L}{4E}\right)\,,\nonumber\\
P_{\nu_\alpha\rightarrow\nu_\alpha}&=&1-\sin^2 2 
\theta\sin^2\left(\frac{\dm{}L}{4E}\right)\,,
\end{eqnarray}
where $L$ is the distance traveled by the neutrinos and is usually called
baseline and $E$ is the neutrino energy.
$P_{\nu_\alpha\rightarrow\nu_\beta}$ is called 
appearance probability, since the 
flavor $\beta$ appears as final state and analogously 
$P_{\nu_\alpha\rightarrow\nu_\alpha}$ is called disappearance probability, since
the flavor $\alpha$ disappears. Obviously the two probabilities fulfill the
unitarity condition $P_{\nu_\alpha\rightarrow\nu_\beta}+P_{\nu_\alpha\rightarrow\nu_\alpha}=1$.
Moreover $P_{\nu_\alpha\rightarrow\nu_\beta}$ is invariant under time reversal and 
\abr{cp}-conjugation, since in the two neutrino case there is no \cpv\ in
neutrino oscillations for the same reason as there would be no \cpv\
in the quark sector if only two families existed~\cite{Kobayashi:1973fv}.

\begin{figure}[tb!]
\begin{center}
\includegraphics[angle=-90,width=\textwidth]{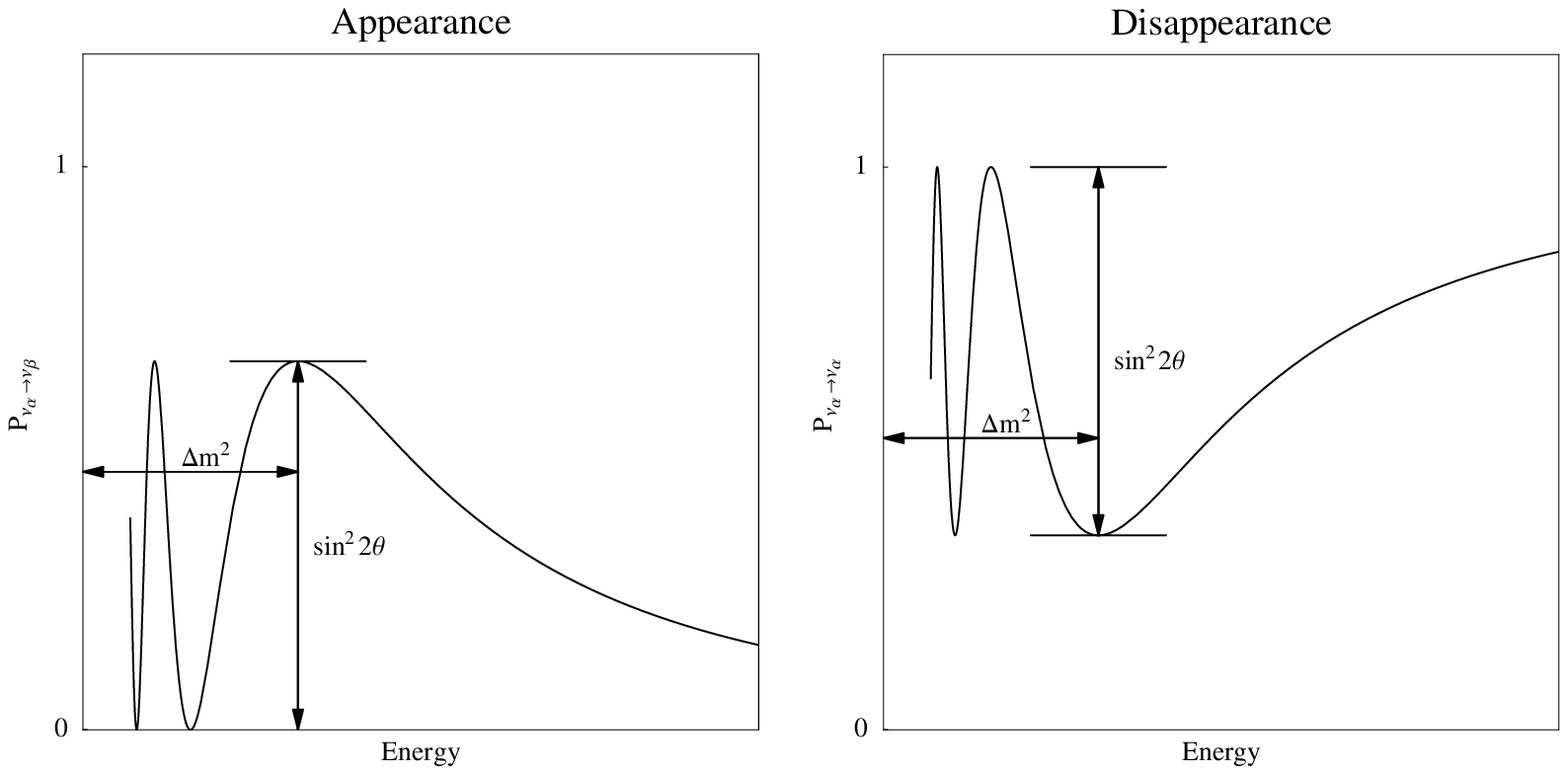}
\end{center}
\mycaption{\label{fig:dip} 
The oscillation probability as a function of the energy in arbitrary 
units. The left hand panel shows the signature of the mixing angle $\theta$
(vertical arrow) and the one of the mass splitting 
$\dm{}$ (horizontal arrow)
in the case of an appearance experiment, whereas the right hand panel shows the
signatures in the case of a disappearance experiment.}
\end{figure}

The parameters $\dm{}$ and $\theta$ are fundamental constants like the 
electron mass or the Cabibbo-angle. However the baseline and neutrino energy
can in principle be chosen by the experimental setup. The signature for the
value of the mixing angle in an 
appearance experiment, \ie\ an experiment which
observes $P_{\nu_\alpha\rightarrow\nu_\beta}$, is given by the height 
of the oscillation peak, which is also indicated by the vertical arrow
 in the left hand panel of \fig~\ref{fig:dip}.
The value of $\dm{}$ is given by the position of the oscillation peak as a
function of the energy, which is shown as horizontal arrow. For a disappearance
experiment the oscillation peak becomes an oscillation dip as shown in the
right hand panel of \fig~\ref{fig:dip}. The depth of the dip is now the
signature for the mixing angle as indicated by the vertical arrow. The position
of the dip yields the value of the mass splitting and is indicated by the
horizontal arrow.

 For both kinds of experiments, appearance and disappearance, there can 
be  a correlation between the measured values of $\dm{}$ and
$\theta$, \ie\ an error on the determination of one parameter introduces an
additional uncertainty on the other parameter. Furthermore an experiment needs
to have enough energy resolution to clearly determine the position of the
peak, otherwise the experiment sees an energy independent signal proportional
to $1/2\sin^22\theta$. Another important factor for the determination of the
mass splitting is the energy calibration of the detector -- any error on the
absolute energy scale directly translates into an error in the position
of the oscillation peak or dip. The major difference between the two possible
experiments is that an appearance experiment is much more sensitive to small
values of $\theta$, because the measurement is performed relative to zero, 
whereas a disappearance experiment measures relative to unity. This implies 
a different behavior of the two types of experiments with respect to certain
systematical errors. On the one hand, the level of background
is crucial for an appearance experiment, since a large background reduces 
the sensitivity to small values of $\theta$. On the other hand, the total 
normalization is vital for a disappearance measurement, because a large
normalization error makes it impossible to detect deviations from unity.

\begin{figure}
  \centerline{\epsfig{file=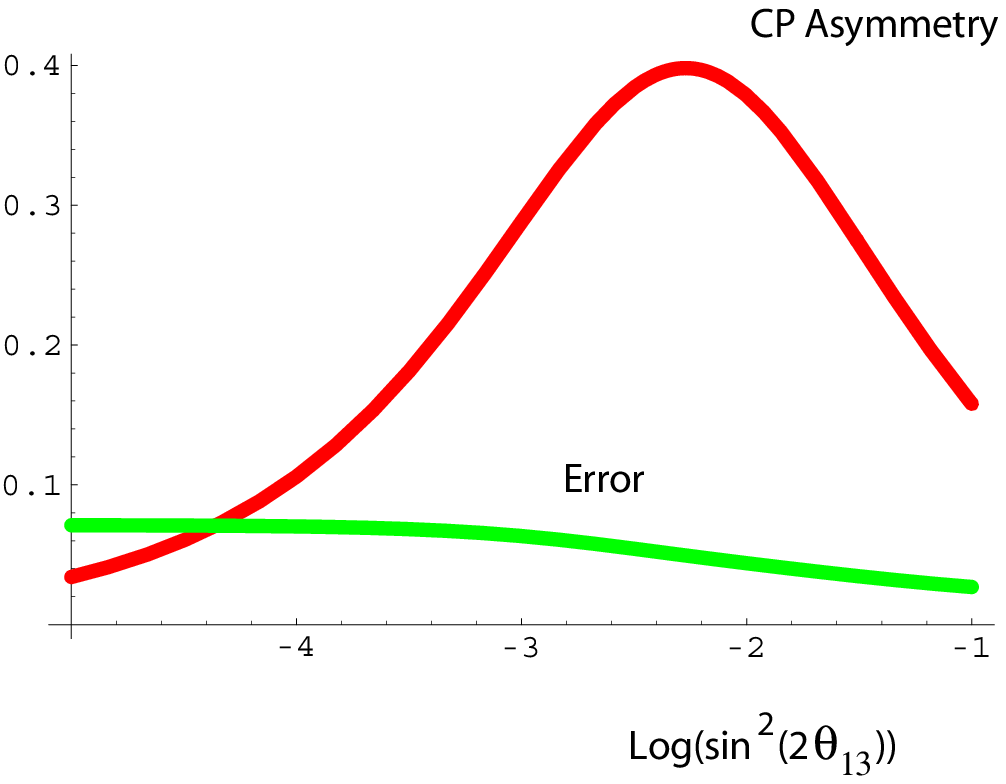, width=0.50\textwidth}}
\mycaption{
Magnitude of the CP asymmetry at the first oscillation maximum,
 for $\delta=1$ as a function of the mixing angle $\sin^2{2\thetaot}$.
 The curve marked 'error'
 indicates the dependence of the statistical+systematic error on such a measurement.
 The curves have been computed for the baseline Beta Beam option at
 the fixed energy $E_\nu=0.4$ GeV, L=130 km, statistical + 2\% systematic
 errors.  }
  \label{fig:Asymm}
\end{figure}

In the full three flavor case, like in the quark sector mixing can
cause CP violation
\begin{equation}
P(\nu_\alpha \rightarrow \nu_\beta)-P(\bar\nu_\alpha \rightarrow
\bar\nu_\beta) \neq 0
\end{equation}
The size of this effect is proportional to
\begin{equation}
J_{CP}=\frac{1}{8}\cos\theta_{13}\sin2\theta_{13}\sin2\theta_{23}\sin2\theta_{12}\sin\delta
\end{equation}
The experimentally most suitable transition to study CP violation
is $\nu_e\leftrightarrow\nu_\mu$, basically because there are
techniques to produce beams of $\nu_\mu$ or $\nu_e$ as well as
detectors for them. In any case energies above the muon threshold are
needed, which are only available in beam experiments. A common tool to
gain some insight into how CP effects are measured is the use of the
so called CP asymmetry asymmetry $A_{CP}$:
\begin{equation}
 A_{CP}=\frac{P(\numunue)-P(\nubarmunubare)}{ P(\numunue)+P(\nubarmunubare)}
\end{equation}
displayed in Fig.~\ref{fig:Asymm},
 or the equivalent time reversal asymmetry $A_T$.

The asymmetry can be large and its value increases for decreasing values of \thetaot
 up to the value when the two oscillations (solar and atmospheric) are of the 
same magnitude. The following remarks can be made:
\begin{enumerate}
        \item
The ratio of the asymmetry to the statistical error is fairly independent on
 \thetaot for large values of this parameter, which explains the relative flatness 
 of the sensitivity curves. 
        \item
This asymmetry is valid for the first maximum. At the second oscillation maximum the 
curve is shifted to higher values of \thetaot so that it could be then an interesting
 possibility for measuring the CP asymmetry, although the reduction in flux is considerable
 (roughly factor 9).  
        \item
The asymmetry has opposite sign for \nuenumu   and \nuenutau, and changes sign when
 going from one oscillation maximum to the next. 
        \item
The asymmetry is small for large values of \thetaot placing a challenging emphasis on
 systematics.
\end{enumerate}

In many cases the propagation of neutrinos does not take place in
vacuo but in matter. Although the interaction of neutrinos with matter
is tiny, matter can have a substantial impact on the oscillation
probabilities.  The weak interaction couples the neutrinos to matter
and besides hard scattering events there is also coherent forward
scattering in very much the same fashion as for visible light
traveling through glass. The point is that the coherent forward
scattering amplitudes are not the same for all neutrino flavors,
since ordinary matter is made of particles of the first family and
does specifically not contain muons or tau-leptons. All flavors have
the same amplitude for neutral current reactions but the electron
neutrinos have an additional contribution due to charged current
reactions. The electron (anti-)neutrino is the only one which can
scatter coherently with the electrons in the matter via the charged
current and this yields an additional contribution to the potential
$A$ for electron (anti-)neutrinos of
\begin{equation}
A=(-)2\sqrt{2}\,G_F\, n_e\, E\,,
\end{equation}
where $G_F$ is the Fermi coupling constant, $n_e$ is the electron
density and $E$ is the neutrino energy. The minus sign is for
anti-neutrinos. In matter the Schr\"odinger equation for neutrino
propagation is now modified by a term containing the potential $A$.
This potential gives rise to an additional phase for $\nu_e$ and thus
changes the oscillation probability. This has two consequences:
\begin{equation}
A_{CP} \neq 0
\end{equation}
even if $\delta=0$, since the potential distinguishes neutrinos
from anti-neutrinos. The second consequence of the matter potential is
that there can be a resonant conversion -- the MSW effect. The
condition for the resonance is
\begin{equation}
\dm{} \simeq A 
\end{equation}
Obviously the occurrence of this resonance depends on the signs of both
sides in this equation. Thus oscillation becomes sensitive to the mass
ordering
\begin{center}
\begin{tabular}{c|cc}
&$\nu$&$\bar\nu$\\
\hline
$\dm{}>0$&MSW&-\\
$\dm{}<0$&-&MSW
\end{tabular}
\end{center}

The general, exact expression for the three flavor oscillation
probabilities in matter~\cite{Ohlsson:1999xb} is rather long and does
therefore not provide much insight. Without going into the details it
is noteworthy that the $\nu_e$ to $\nu_\mu$ transition has the richest
structure in terms of effects which in principle can be extracted.
This, however, also implies that there are strong correlations,
especially between $\deltacp$ and $\theta_{13}$~\cite{Cervera:2000kp}. Moreover, there can
be up to eight discrete, degenerate solutions~\cite{eight}. This
problem has been widely recognized and a large number of solutions
have been proposed like including additional final states, \eg\
$\nu_\tau$ or to use different energies and baselines. 


\subsection{ Neutrino mass limits from laboratories}

Direct laboratory limits on neutrino masses are obtained from
kinematical studies.
The most stringent current upper limit is that on the $\bar{\nu}_e$ mass, 
coming from the Mainz experiment measuring the end-point of the electron energy spectrum in
Tritium beta decay~\cite{Weinheimer:2003fj}
$$
m_{\nubare} \leq 2.2  ~{\rm eV} \;{(\rm 95\%CL)}
$$
The Troitsk group has also published a similar limit 
\cite{Troitsk}:
$$
m_{\nubare} \leq 2.1  ~{\rm eV} \;{(\rm 95\%CL)}
$$
however they must include an ad-hoc step function near the endpoint to avoid the
problem of negative mass squared.

The proposed
KATRIN experiment aims to improve the sensitivity to 
$m_{\bar {\nu_{e}}} \sim 0.3$~eV \cite{KATRIN}.
Similar sensitivities are the goal of the longer term MARE
experiment \cite{Mare} based on an array of several thousand
of microbolometers.
These measurements are sensitive to:
\begin{equation}
m_{\bar \nu_e} =
\bigg(\sum_i |U_{ei}^2|\ m_i^2\bigg)^{1/2} =
\bigg(\cos^2 \theta_{13}(m_1^2 \cos^2\theta_{12} +
m_2^2  sin^2{\theta_{12}}) + m_3^2 sin^2{\theta_{13}}\bigg)^{1/2}
\end{equation}

Limits to neutrino masses come also from cosmology \cite{Fukugita},
combining results from cosmic microwave anysotropies, supernovae surveys,
galaxy clustering and Lyman $\alpha$ cloud absorption power,
limits on the sum of the neutrino masses of the order of 1 eV can be 
derived.

An important constraint on Majorana neutrino masses 
arises from neutrinoless double-$\beta$
decay, in which an $(A,Z)$ nucleus decays to $(A,Z+2)
+ 2 \ e^-$, without any neutrino emission. 
This process can be used to constrain the combination
\begin{equation}
|m_{\beta\beta}| = \bigg|\sum_i U_{ei}^{*2}\ m_i \bigg| =
\bigg|\cos^2 \theta_{13}(m_1
\cos^2\theta_{12} +  m_2 e^{2i\alpha} sin^2\theta_{12}) + m_3 e^{2i\beta}
sin^2\theta_{13}\bigg|.
\label{dbeta}
\end{equation}
which involves a coherent sum over all the different
Majorana neutrino masses $m_i$, weighted by their
mixings with the electron flavour eigenstate, which may 
include CP-violating phases, as discussed below. This 
observable is
therefore distinct from the quantity observed in Tritium 
$\beta$ decay. 

The interpretation of neutrinoless
double-$\beta$ decay data depends on calculations of the  
nuclear matrix elements entering in this process.

A claim for a neutrinoless double-$\beta$ signal has been made by \cite{KK}
analyzing the Heidelberg-Moscow data on $^{76}$Ge:
$$
 T^{0\nu}_{1/2}=1.19 \cdot 10^{25} \; {\rm years}
$$
corresponding to
$$
<m_{\beta\beta}>=0.05-0.85\;{\rm eV} \;{\rm (95\%CL)}
$$
the uncertitude coming from the choice of the nuclear matrix element calculation.

This result is in contrast with the limit computed with a combined analysis
of a subset of the Heidelberg-Moscow data and IGEX experiments \cite{HMIgex} and
 to what reported by a separate group of the original 
collaboration \cite{HM}, reporting no evidence for a signal.

Recent results on $^{130}Te$ from the Cuoricino collaboration \cite{Cuoricino}:
$T^{0\nu}_{1/2}>1.8\cdot10^{24}\;{\rm years}$ corresponding to 
$m_{\beta\beta}<.2-1.1\;{\rm eV}$
and on $^{100}Mo$ from the NEMO3 collaboration \cite{Nemo3}:
$T^{0\nu}_{1/2}>4.6\cdot10^{23}\;{\rm years}$ corresponding to 
$m_{\beta\beta}<.7-2.8\;{\rm eV}$ do not confirm the Germanium claim,
but are not sensitive enough to rule out it.

The approved future experiments at LNGS CUORE \cite{Cuore} and GERDA \cite{Gerda}
 will have
the required sensitivity to unambiguously clarify this experimental
situation: having a sensitivity of $m_{\beta\beta}=0.024-0.14$ eV and $m_{\beta\beta}=0.09-0.29$ eV
respectively.

\section{Description of the accelerator neutrino facilities}

\subsection{Present generation of long-baseline experiments}

\footnote{Material for this Section is mainly taken from ref.~\cite{OurReview}}
 Over the next five years the present generation of oscillation experiments at
 accelerators with long-baseline $\nu_{\mu}$ beams (Table~\ref{tab:beams1}),
 K2K at KEK \cite{K2K}, MINOS \cite{Minos}  at the NuMI beam from FNAL \cite{NUMI} and ICARUS \cite{ICARUS} and OPERA
 \cite{OPERA} at the CNGS beam from CERN \cite{CNGS} are expected
 to confirm the
 atmospheric evidence of oscillations and measure $\sin^2 2 \theta_{23}$ and
 $|\Delta m^2_{23}|$ within $10 \div 15 $ \%  of accuracy if
 $|\Delta m^2_{23}| >  10^{-3}$ eV$^2$.
 K2K and MINOS are looking for neutrino disappearance, by measuring
 the $\nu_{\mu}$ survival probability as a function of neutrino energy while
 ICARUS and OPERA will search for evidence
 of $\nu_{\tau}$ interactions in a  $\nu_{\mu}$ beam, the final
 proof of \numunutau oscillations.
 K2K has already completed its data taking at the end of 2004, while
 MINOS has started data taking beginning 2005. CNGS is expected to start
 operations in the second half of 2006.

 \begin{table*}[htb]
 \mycaption{Main parameters for present long-baseline neutrino beams}
 \begin{tabular}{lccccc}
 \hline
  Neutrino facility  &Proton momentum (GeV/$c$)&L (km)& $E_{\nu}$
 (GeV) & pot/yr ($10^{19}$)\\
 \hline
              KEK PS       &   12    & 250    &   1.5      &  2   \\
              FNAL NuMI    &  120    & 735    &   3        &  20$\div$ 34\\
              CERN CNGS    &  400    & 732    &   17.4     &  4.5$\div$ 7.6\\
 \hline
 \end{tabular}
 \label{tab:beams1}
 \end{table*}

 In all these facilities conventional muon neutrino beams are produced through the
 decay of $\pi$ and K mesons generated by a high energy proton beam hitting needle-shaped
 light targets. Positive (negative) mesons are sign-selected and focused (defocused)
 by large acceptance  magnetic lenses into a long evacuated decay
 tunnel where $\nu_{\mu}$'s  ($\overline{\nu}_{\mu}$'s) are generated.
 In case of positive charge selection, the $\nu_{\mu}$ beam has typically
 a contamination of $\overline{\nu}_{\mu}$ at few percent level
 (from the decay of the residual $\pi^{-}, K^{-}$ and $K^0$) and
 $\sim 1 \%$ of  $\nu_e$ and $\overline{\nu}_e$
 coming from three-body $K^{\pm}$, $K_0$ decays and $\mu$ decays.
 The precision on the evaluation of the intrinsic $\nu_e$ to $\nu_\mu$ contamination  is
 limited by the knowledge of the $\pi$ and $K$ production in the primary proton beam target.
 Hadroproduction measurements at 400 and 450 GeV/c performed with the NA20 \cite{NA20} and SPY
 \cite{SPY}
 experiments at the CERN SPS provided results with  $5 \div 7 \%$ intrinsic systematic uncertainties.

  The CNGS $\nu_\mu$  beam has been optimized for the $\nu_\mu \rightarrow \nu_\tau$
 appearance search.
 The beam-line design was accomplished
 on the basis of the previous experience with the WANF beam at CERN SPS \cite{WANF}.
 The expected muon neutrino flux at the Gran Sasso site will have
 an average energy of $17.4$ GeV and $\sim 0.6 \%$ $\nu_e$ contamination
 for $E_\nu < 40$ GeV. Due to the long-baseline  (L=732 Km) the contribution
 to neutrino beam from the $K^0$ and mesons produced in the reinteraction processes
 will be  strongly reduced with respect to the WANF \cite{WANF1}: the $\nu_e/\nu_{\mu}$ ratio
  is expected to be known within  $\sim 3 \%$ systematic uncertainty \cite{cngs_syst}.

 Current long-baseline experiments  with conventional neutrino beams  can look for
 $\nu_{\mu} \rightarrow \nu_e$ even if they are
 not optimized for $\theta_{13}$ studies.
  MINOS at NuMI is expected to reach a sensitivity of $\sin^2{2\thetaot}=0.08$
 \cite{Minos} integrating $14{\cdot}
 10^{20}$ protons on target (pot)
  in 5 years according to the FNAL proton plan evolution \cite{FermilabProtons}.
  MINOS main limitation
 is the poor electron identification efficiency of the detector.

Thanks to the dense ECC structure and the high granularity provided by
 the nuclear emulsions, the
OPERA detector is also suited for electron and γ detection \cite{OPERA}.
 The resolution in measuring the energy of
an electromagnetic shower in the energy range relevant to CNGS is 
approximately constant and is about 20\%.
 Furthermore, the nuclear emulsions are able to measure the number of grains 
associated to each track.
This allows an excellent two tracks separation (better than 1 $\mu$m).
 Therefore, it is possible to disentangle
single-electron tracks from tracks produced by electron pairs coming from
 $\gamma$ conversion in the lead.
 These features are particularly important for the \numunue analysis.
The outstanding position resolution of nuclear emulsions can also be
 used to measure the angle of each charged track with an accuracy of about
 1 mrad. This allows momentum measurement by using the
Multiple Coulomb Scattering with a resolution of about 20\% and the
 reconstruction of kinematical variables
characterizing the event (i.e. the missing transverse momentum at the
 interaction vertex $p^{\rm miss}_T$ and the
transverse momentum of a track with respect to hadronic shower direction $Q_T$).

The expected number of events for the \numunue oscillation search is reported
in Table \ref{Table:Operaevents}

\begin{table}[hbtp]
  \small
\begin{center}
\mycaption{Expected number of signal and background events and
analysis efficiencies $\epsilon$ for
 OPERA \cite{Komatsu, Migliozzi} assuming 5 years data taking  with the nominal CNGS beam and oscillation
parameters $\Delta m_{23}^2 =2.5\times10^{-3}~\mbox{eV}^2$,
$\theta_{23}=45^\circ$ and $\theta_{13}=5^\circ$.}
\vspace{5mm}
\begin{tabular}{||c|c|c|c|c|c|c||}
\hline
$\theta_{13}$ & $\sin^22\theta_{13}$ & $\nu_e$CC signal & $\tau\rightarrow e$ & $\nu_\mu\mbox{CC}\rightarrow\nu_\mu\mbox{NC}$ & $\nu_\mu$NC & $\nu_e$CC beam  \\
\hline
$7^\circ$ & 0.058 & 5.8 & 4.6 & 1.0&5.2& 18 \\ 
\hline
$\epsilon$ &   &0.31 & 0.032 & $3.4\cdot 10^{-5}$ & $7\cdot 10^{-5}$&0.082 \\
\hline
\end{tabular}
\label{Table:Operaevents}
\end{center}
\end{table}

 OPERA  can reach a 90\% C.L. sensitivity
  $\sin^2 2 \theta_{13}=0.06$ 
 ($\Delta m^2_{23} = 2.5 {\cdot} 10^{-3}$  eV$^2$, convoluted to CP and matter effects) \cite{Komatsu,Migliozzi},
 a factor $\sim 2$ better than Chooz
 for five years exposure to the
 CNGS beam at nominal intensity for shared operation $4.5 {\cdot} 10^{19}$ pot/yr.

\begin{figure}
  \begin{center}
  \mbox{\epsfig{file=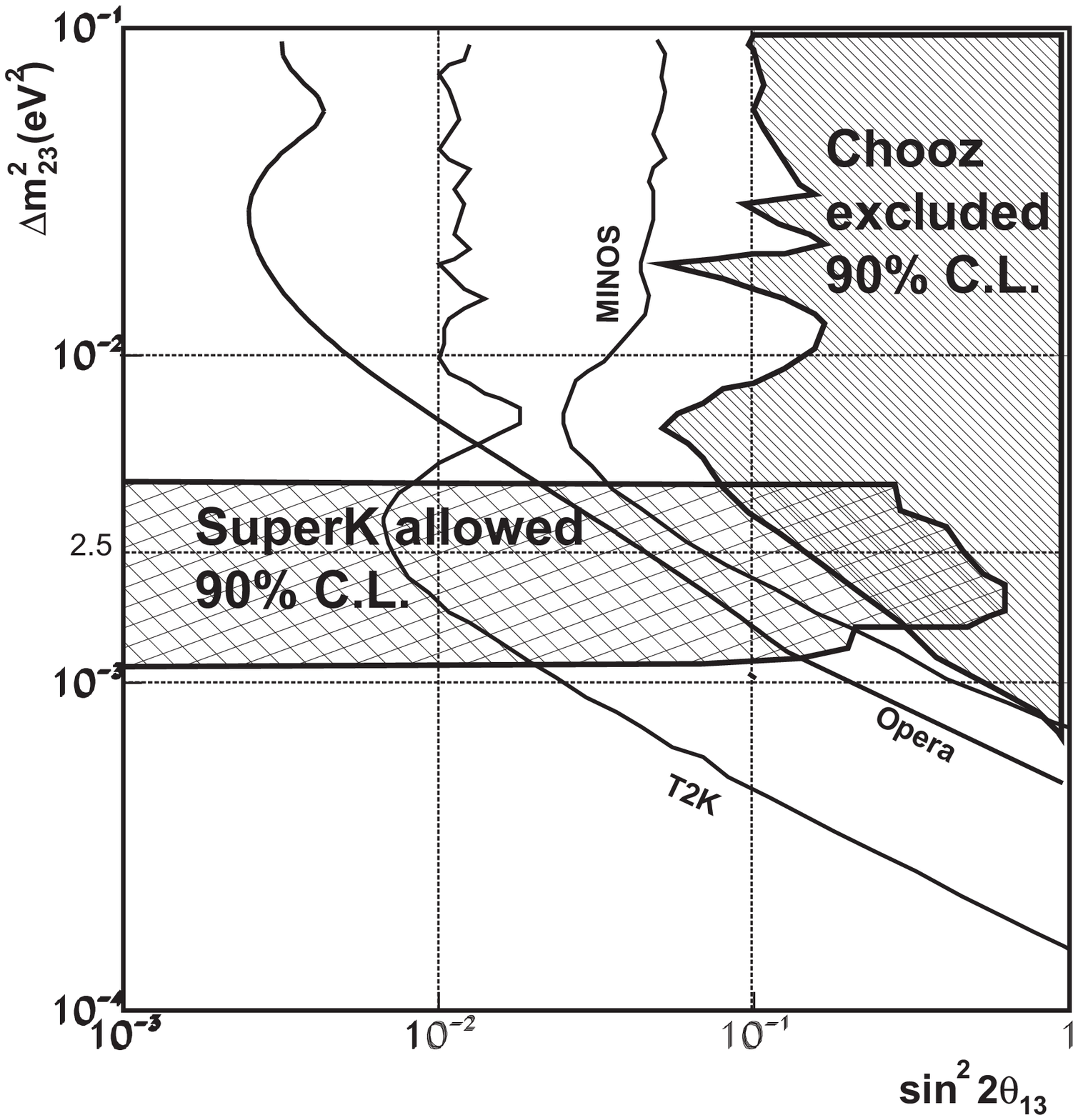, width=8.0cm, height=8cm}}
  \end{center}
  \vskip -0.5cm
  \mycaption{ Expected sensitivity on $\theta_{13}$ mixing angle
               (matter effects and  CP violation effects not included)
               for MINOS, OPERA and for the next T2K experiment,
               compared to the Chooz exclusion plot, from \cite{OurReview}.}
  \label{exclusion}
 \end{figure}

 A plot of \thetaot sensitivities is reported in Fig.~\ref{exclusion}.
 According to  the CERN  PS and SPS upgrade studies
 \cite{SPS_pot_increase},
 the CNGS beam intensity could be improved by a factor 1.5, allowing for
  more sensitive neutrino  oscillation searches for the OPERA experiment.

  It is worth mentioning that
  the sensitivity on $\theta_{13}$ measurement of the current long-baseline experiments with
 conventional neutrino beams, like NuMI and CNGS, will be limited by the power of the proton source
 which determines the neutrino flux and the event statistics, by the not optimized $L/E_\nu$
 and by the presence of the $\nu_e$ intrinsic beam contamination and
 its related systematics.
 This is particular true for CNGS where the neutrino energy,
 optimized to overcome the kinematic
 threshold for $\tau$ production and to detect the $\tau$ decay products, s about
 ten times higher the optimal value for \thetaot searches at that baseline.

 Another approach to search for non vanishing \thetaot is to look at \nubare\  disappearance
 using nuclear reactors as neutrino source.

 The Double-Chooz experiment aims at improving the current knowledge on
 $\theta_{13}$ by observing the disappearance of $\bar\nu_e$ from
 nuclear reactors.  The relevant oscillation probability is
 \begin{equation}
 \label{eq:pee}
 P(\bar\nu_e\rightarrow\bar\nu_e)\simeq 1 -
 \sin^22\theta_{13}\sin^2\left(\frac{\Delta m^2_{31}L}{4E}\right)+\ldots
 \end{equation}
 which does not depend on $\theta_{23}$ and the CP-phase $\delta_{CP}$.
 The dependence on $\Delta m^2_{21}$ and $\theta_{12}$ is negligible
 for the chosen baseline. Therefore this approach allows a unambiguous
 detection of $\theta_{13}$ free of correlations and degeneracies.  As
 it is obvious from eq.~\ref{eq:pee} the measurement requires a very
 precise control of the absolute flux. For that reason Double-Chooz
 will employ a near and far detector. The direct comparison of the
 event rates in each detector will allow to cancel many systematical
 errors and thus is essential in reaching the required low level of
 residual errors. Both detectors need some overburden to reduce the
 cosmic muon flux to an acceptable level.  The advantage of
 Double-Chooz is that it will use an existing cavern for the far
 detector, which puts it ahead in time of any other reactor experiment,
 provided that the final funding decision is made in a timely manner.
 \begin{table}[h]
 \begin{center}
 \mycaption{\label{choozto2chooz} Overview of the systematic errors of
   the CHOOZ and Double-CHOOZ experiment. Table taken from~\cite{DChooz}}
 \begin{tabular}{lrr}
 \hline
  & \multicolumn{1}{c}{CHOOZ} & \multicolumn{1}{c}{Double-CHOOZ}  \\
 \hline
  Reactor cross section & 1.9~\% & ---    \\
  Number of protons     & 0.8~\% & 0.2~\%  \\
  Detector efficiency   & 1.5~\% & 0.5~\%  \\
  Reactor power         & 0.7~\% & ---    \\
  Energy per fission    & 0.6~\% & ---    \\
 \hline
 \end{tabular}
 \end{center}
 \end{table}
 This cavern is $1.05\,\mathrm{km}$ away from the reactor and the near
 detector will be at $\sim 200\,\mathrm{m}$ from the reactor.  Both
 detectors will be based on a Gadolinium loaded liquid scintillator and
 use inverse $\beta$-decay and the delayed neutron capture signal. Both
 detector will have a fiducial mass of $10.16\,\mathrm{t}$. The
 improvements in systematical accuracy need are summarized in
 table~\ref{choozto2chooz}.

 The sensitivity after 5 years of data taking will be
 $\sin^22\theta_{13}=0.02$ at $90$\% CL~\cite{DChooz}, which could be achieved
 as early as 2012. It is conceivable to use a larger, second cavern to
 place a $200\,\mathrm{t}$ detector to
 even improve that bound down to $\sin^22\theta_{13}<0.01$~\cite{Huber:2006vr}.

\begin{figure}
\centerline{\epsfig{file=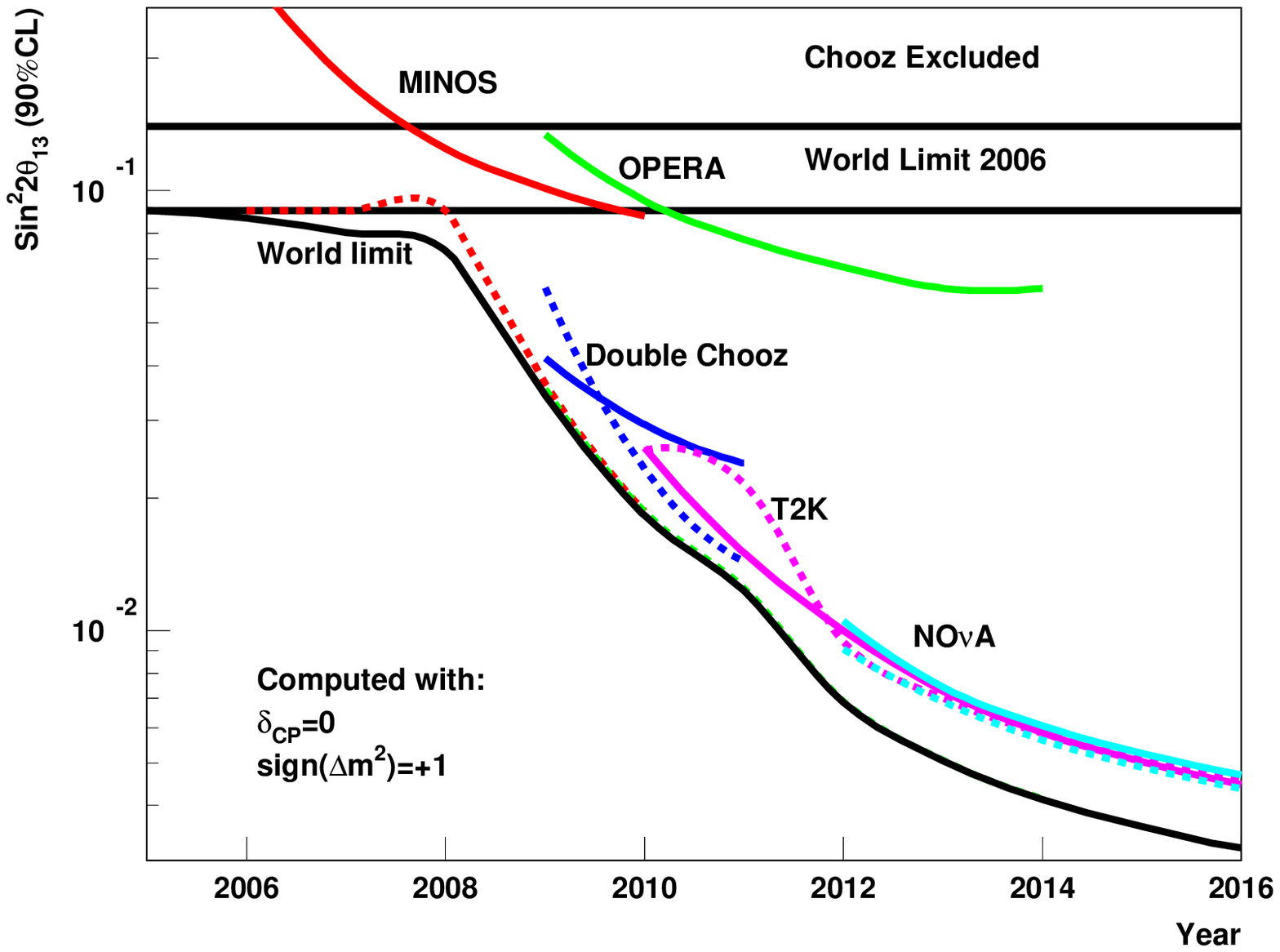,width=0.70\textwidth}}
    \mycaption{Evolution of sensitivities on $\sin^2{2\thetaot}$
    as function of time. For each experiment are displayed the sensitivity as
    function of time (solid line) and the world sensitivity computed without
    the experiment (dashed line). The comparison of the two curves shows the discovery potential
    of the experiment  along its data taking. The world overall sensitivity along the
    time is also displayed. The comparison of the overall world sensitivity
    with the world sensitivity computed without a single experiment
    shows the impact of the results of the single experiment.
    Experiments are assumed to provide results
    after the first year of data taking.}
    \label{fig:th13vstime}
\end{figure}

 A sketch of \thetaot sensitivities as a function
 of the time, following the schedule reported in the experimental proposals, computed
for the approved experiments, is
 reported in Fig.~\ref{fig:th13vstime}.

  According to the present experimental situation, conventional neutrino beams can be improved
  and optimized for the $\nu_{\mu} \rightarrow \nu_e$ searches.
  The design of a such new SuperBeam facility for a  very high intensity and low energy
  $\nu_\mu$ flux will demand:

 \begin{itemize}
   \item a new  higher power proton driver, exceeding the megawatt,
         to deliver more intense proton beams on target;
   \item a tunable $L/E_{\nu}$ in order to explore the $\Delta m^2_{23}$
         parameter region as indicated by the previous experiments with neutrino
         beams and atmospheric neutrinos;
   \item narrow band beams with $E_{\nu} \sim 1\div 2$ GeV;
   \item a lower intrinsic $\nu_e$ beam contamination which can be obtained
         suppressing the $K^+$ and $K^0$
         production by the primary proton beam in the target.
 \end{itemize}

 \noindent An interesting option for  the SuperBeams is the possibility to tilt the beam
 axis a few degrees with respect to the position of the far detector (Off-Axis beams)
 \cite{T2K, OffAxis}.
 According to the two body $\pi$-decay kinematics,   all the pions above a given
 momentum
 produce neutrinos of similar energy at a given angle $\theta \ne 0$
 with respect to the direction of parent pion (contrary to the $\theta=0$ case where
 the neutrino energy is proportional to the pion momentum).
 \noindent These neutrino beams have several advantages with respect to the
 corresponding on-axis ones:
 they are narrower, lower energy and with a smaller \nue\
 contamination (since \nue\  mainly come from three body decays)
 although the neutrino flux can be significantly smaller.

  The intrinsic limitations of conventional neutrino beams are overcome
 if the neutrino parents can be fully selected, collimated and accelerated to a given energy.
 This  can be attempted within the muon or a beta decaying ion lifetimes.
 The neutrino beams from their decays would then be pure and perfectly predictable.
 The first approach brings to the Neutrino Factories \cite{Nufact},
 the second to the BetaBeams \cite{BetaBeam}.
 However, the technical difficulties associated
 with developing and building
 these novel conception neutrino beams  suggest for the middle
 term option to improve  the conventional beams by new high intensity
 proton machines,
 optimizing the beams for the $\nu_\mu \rightarrow \nu_e$ oscillation
 searches (SuperBeams).

\subsection{
\label{sec:oa} Off axis SuperBeams: T2K, T2HK and NO\boldmath{$\nu$}A}
The T2K (Tokai to Kamioka) experiment \cite{T2K} will aim neutrinos from the Tokai site to the
 Super-Kamiokande detector 295 km away.
The neutrino beam is produced by pion decay from a horn focused beam, with a system of three horns and reflectors. 
The decay tunnel length (130 m long) is optimized for the decay of 2-8 GeV pions and
 short enough to minimize the occurrence of muon decays. 
The neutrino beam is situated at an angle of 2-3 degrees from the direction of the Super-Kamiokande detector, assuring a pion decay peak energy of 0.6 GeV.
 The beam line is equipped with a set of dedicated on-axis and off-axis near detectors at
 the distance of  280 meters.
\begin{figure}
  {\epsfig{file=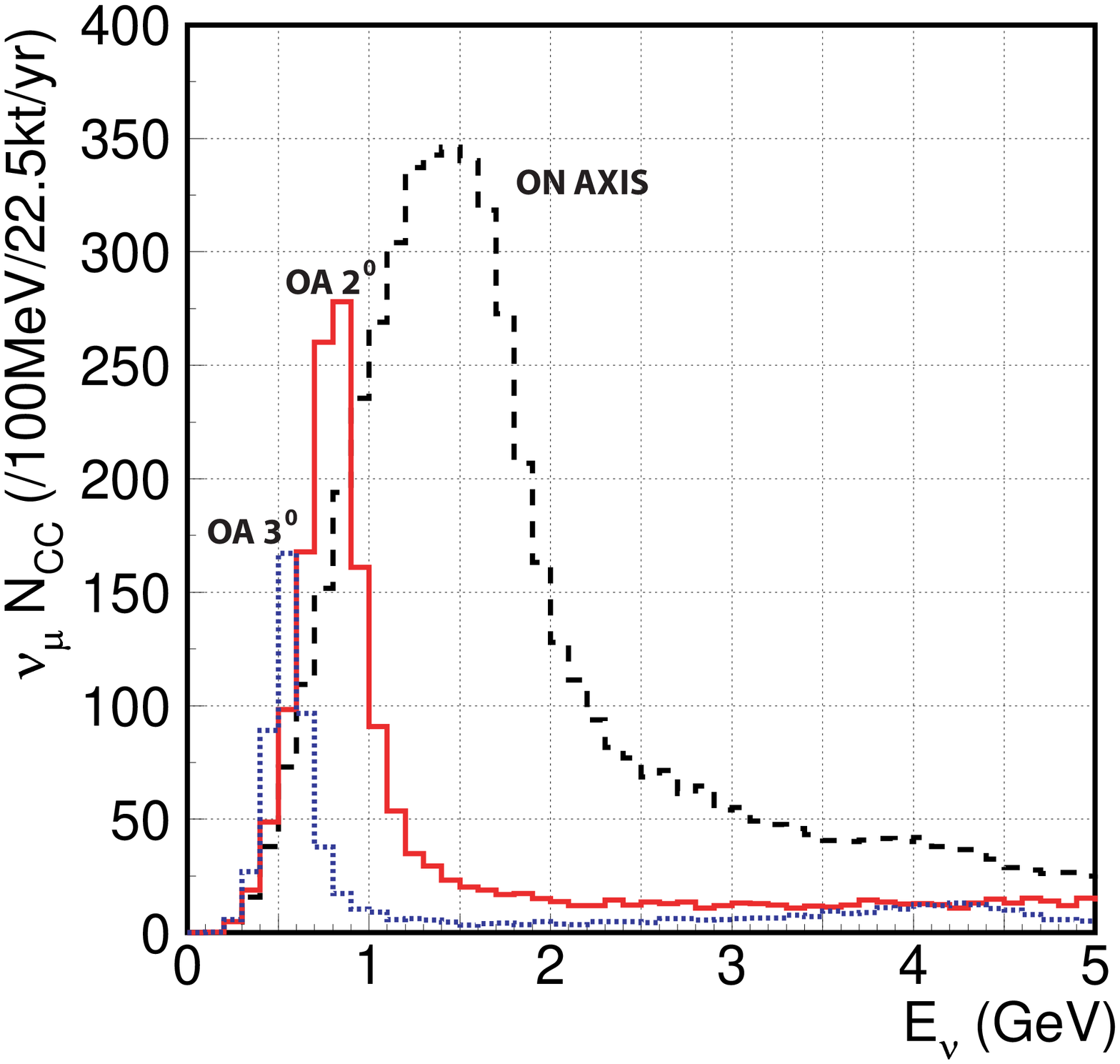, width=0.36\textwidth}} \hfill
  {\epsfig{file=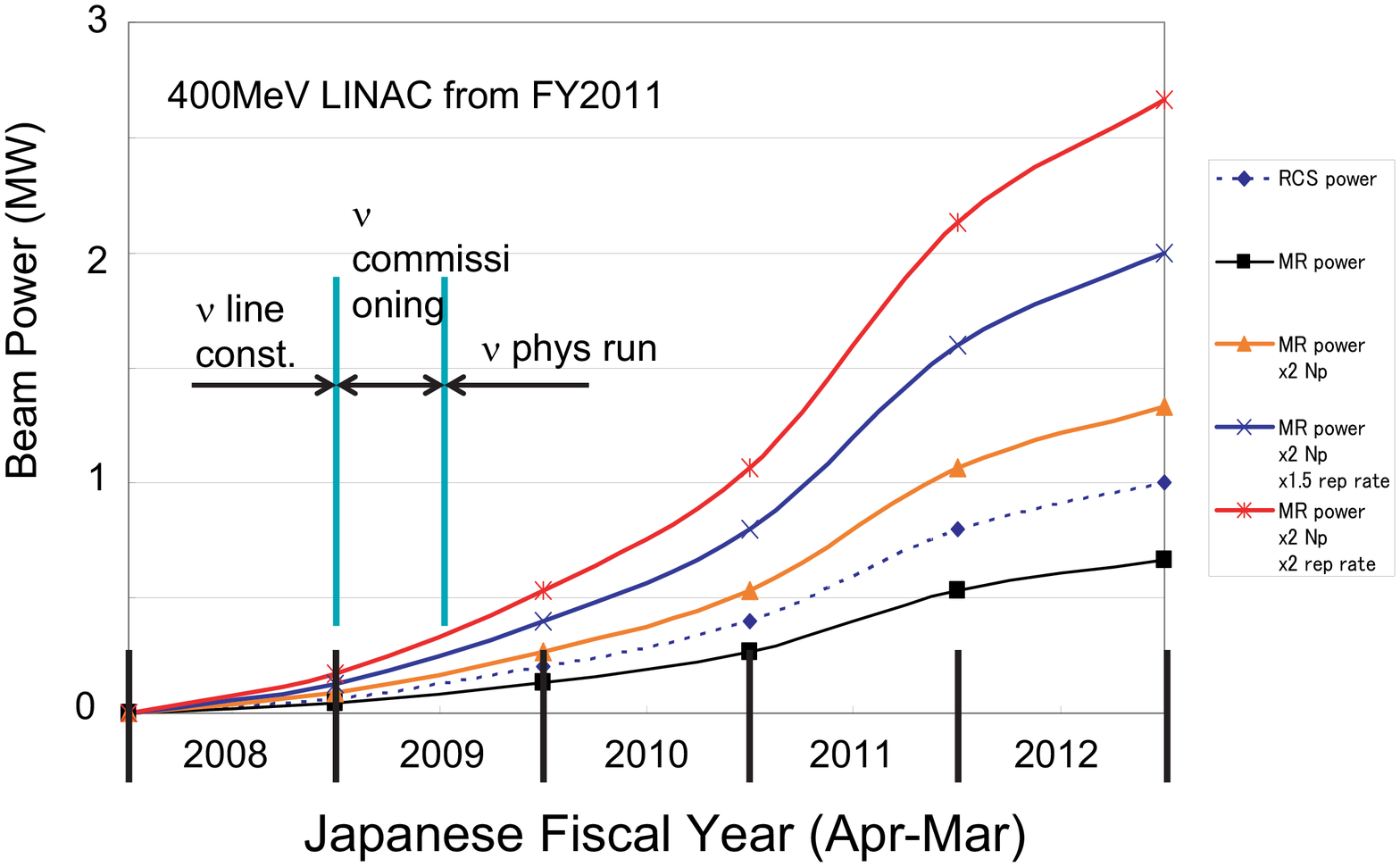, width=0.55\textwidth}}
  \vskip -0.5cm
  \mycaption{Left: T2K neutrino beam energy spectrum for different
               off-axis angle $\theta$. Right: expected evolution of T2K
      beam power as function of time. Baseline option is the second lowest solid curve.}
  \label{off-axisJHF}
 \end{figure}

The main goals of the experiment are as follows:
\begin{enumerate}
	\item
The highest priority  is the search for $\nu_e$ appearance to detect sub-leading
\numunue\ oscillations.
  It is expected that the sensitivity of the experiment in a 5 years $\nu_\mu$ run,
 will be of the order of $\sin^2{2\thetaot} \leq 0.006$ \cite{T2K}. 
	\item
Disappearance measurements of $\nu_\mu$.
 This will  improve measurement of $\Delta m^2_{23}$ down to a precision of a 0.0001 eV$^2$ or so.
 The exact measurement of the maximum disappearance is a precise measurement
 of $\sin^2{2\thetatt}$.
  These precision measurements of already known quantities require
 good knowledge of flux shape, absolute energy scale,  experimental energy
 resolution and of the cross-section as a function of energy. 
They will be crucial to measure the tiny \numunue\ oscillations
 \cite{Donini05, Donini:2005db}.
	\item
 Neutral current disappearance (in events tagged by $\pi^\circ$ production) will 
 allow for a sensitive search of sterile neutrino production.
\end{enumerate}

The T2K experiment is planned to start in 2009 with a beam intensity reaching
 1 MW beam power on target after a couple years, see Fig.~\ref{off-axisJHF}.
  It has an upgrade path which involves:
 a further near detector station at 2 km featuring a water \v{C}erenkov detector,
 a muon monitor and a fine grain detector (possibly liquid argon).
 The phase II of the experiment, often called T2HK, foresees
 an increase of beam power up to the maximum feasible with the accelerator and target
 (4 MW beam power), antineutrino runs, and a very large water \v{C}erenkov (HyperKamiokande) with
 a rich physics programme in proton decay, atmospheric and supernova neutrinos
 and, perhaps, leptonic CP violation, that could be built around in about 15-20
 years from now. \\

 The NO$\nu$A experiment with an upgraded NuMI Off-Axis neutrino beam
 \cite{Nova} ($E_{\nu} \sim 2 $
 GeV and a $\nu_e$ contamination lower than $0.5 \%$) and with a baseline of
  810 Km (12 km Off-Axis), has been recently proposed at FNAL with the aim to explore the
 $\nu_{\mu} \rightarrow \nu_e$ oscillations with a  sensitivity 10 times better
 than MINOS.
 If approved in 2006 the experiment could start data taking in 2011.
 The NuMI target will receive
 a 120 GeV/c proton flux with an expected intensity of $6.5 {\cdot} 10^{20}$ pot/year (
 $2 {\cdot} 10^7$ s/year are
 considered available to NuMI operations while the other beams are normalized to
 $10^7$ s/year).
 The experiment will use a near and a far detector, both  using liquid scintillator
 (TASD detector).
 In a 5 years $\nu_\mu$ run, with 30 kton active mass far detector, a sensitivity on
 $\sin^2 2 \theta_{13}$ slightly better
 than  T2K, as well as a precise measurement of  $|\Delta m_{23}^2|$ and 
 $\sin^2 2 \theta_{23}$, can be achieved.
 NO$\nu$A can also allow to solve the mass hierarchy problem for a limited
 range of the \delCP and \sigdm parameters \cite{Nova}.

 \noindent As a second phase, the new proton driver of
 8 GeV/c and 2 MW, could increase the  NuMI beam
  intensity to $17.2 \div 25.2 {\cdot} 10^{20}$ pot/year,
 allowing to improve the experimental sensitivity by a factor two 
 and to initiate the experimental search for the CP violation.

\subsection{SPL SuperBeam}
 In  the CERN-SPL SuperBeam project  \cite{SPL,SPL-Physics,nufact1}
 the planned 4MW SPL (Superconducting Proton Linac)  would deliver a 2.2  GeV/c
 proton beam,  on a Hg target to generate
 an intense $\pi^+$ ($\pi^-$) beam focused by a suitable
 magnetic horn in a short decay tunnel. As a result   an intense
 $\nu_{\mu}$ beam, will be produced
 mainly via the $\pi$-decay,  $\pi^+ \rightarrow \nu_{\mu} \; \mu^+$ providing a
 flux $\phi \sim 3.6 {\cdot} 10^{11} \nu_{\mu}$/year/m$^2$  at 130 Km
 of distance, and an average energy of 0.27 GeV.
 The $\nu_e$ contamination from $K$ will be suppressed by threshold effects
 and the resulting $\nu_e/\nu_{\mu}$ ratio ($ \sim 0.4 \%$) 
  will be known within  $2\%$ error.
 The use of a near and far detector (the latter at $L = 130$ Km of distance
 in the Frejus area \cite{Mosca}, see Sec.~\ref{sec:Memphys})
 will allow for both $\nu_{\mu}$-disappearance and
 $\nu_{\mu} \rightarrow \nu_e$ appearance studies.
 The physics potential of the 2.2 GeV SPL SuperBeam (SPL-SB)
 with a water \v{C}erenkov far detector with a fiducial mass of 440 kton,  has been extensively
 studied \cite{SPL-Physics}. \\

 New developments show that the potential of the SPL-SB potential could be
 improved by rising the SPL energy to 3.5 GeV \cite{Cazes},
 to produce   more copious secondary mesons
 and to focus them more efficiently. This seems feasible if
 status of the art RF cavities would be used in place of the previously
 foreseen LEP cavities \cite{Garoby-SPL}.

\begin{figure}
  \centerline{\epsfig{file=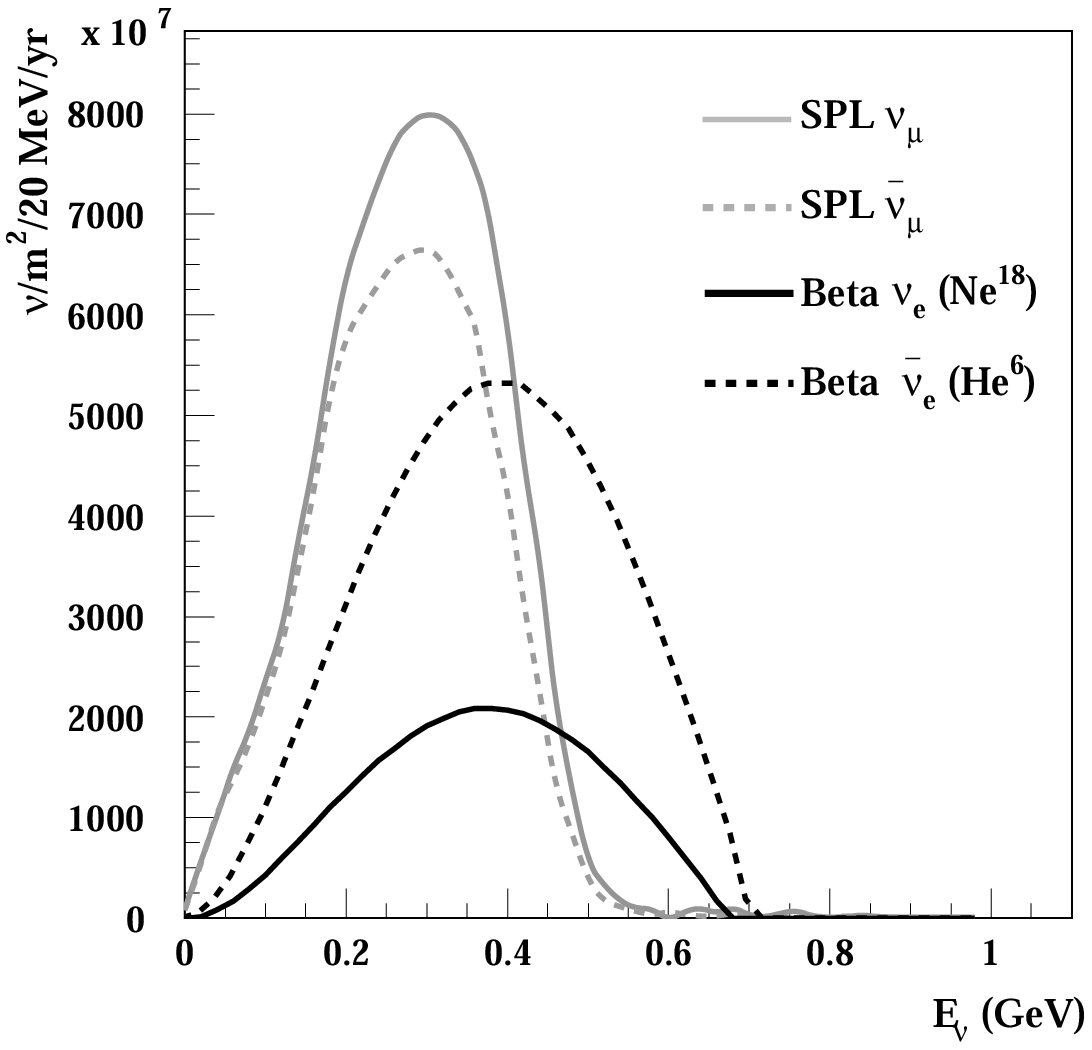,width=0.5\textwidth}}
  \mycaption{Neutrino flux of $\beta$-Beam ($\gamma=100$)
   and CERN-SPL SuperBeam, 3.5 GeV, at 130 Km of distance.}
  \label{fig:fluxes}
 \end{figure}

The focusing system (magnetic horns), originally optimized in the context of a 
Neutrino Factory \cite{SIMONE1,DONEGA},
 has been redesigned considering the specific
 requirements of a Super Beam.
 To obtain a maximum oscillation probability, corresponding to a 
 mean neutrino energy of $300$~MeV,
 one should collect $800$~MeV/c pions, see Fig.~\ref{fig:fluxes}.
At higher beam energy, the kaon rates grow rapidly compared to the pion rates,
 an experimental confirmation
 \cite{Harp} of such numbers would be strongly needed.

 In this upgraded configuration neutrino flux could be increased by a factor $\sim 3$ 
 with respect to the 2.2 GeV configuration, 
the number of expected $\nu_\mu$ charged current is about $95$ per ${\rm kton \cdot yr}$

 A sensitivity $\sin^2{2 \thetaot} < 0.8 \cdot 10^{-3}$ is obtained in a 5 years $\nu_\mu$ plus
5 year \nubarmu\ run ($\delta=0$ intrinsic degeneracy accounted for, sign and octant
  degeneracies not accounted for), allowing
 to discovery CP violation (at 3 $\sigma$ level) if
 $\delCP \geq 25^\circ$  and
 $\theta_{13} \geq 1.4^\circ$ \cite{MMNufact04, Campagne}. The expected
 performances are shown in Fig.~\ref{fig:th13} and \ref{fig:final} along with those
of other setups.

\subsection{BetaBeams}
\label{BetaBeam}

BetaBeams   have been introduced by
P. Zucchelli in 2001 \cite{BetaBeam}.
The idea is to generate pure, well collimated and intense
\nue\  (\nubare) beams by producing, collecting, accelerating radioactive ions
and storing them in a decay ring in 10 ns long bunches, to suppress
the atmospheric neutrino backgrounds.
The resulting BetaBeam  fluxes
could be easily computed by the properties of the beta decay of the parent
ion and by its Lorentz boost factor $\gamma$ and would not be contaminated
by unwanted neutrino flavours or helicities.
 The best ion candidates so far
 are  $^{18}Ne\;$  and $^6He\;$ for \nue\ and \nubare\  respectively.
The schematic layout of a Beta Beam is the following (see also Fig.~\ref{fig:sketch}):
\begin{figure}
 \centerline{\epsfig{file=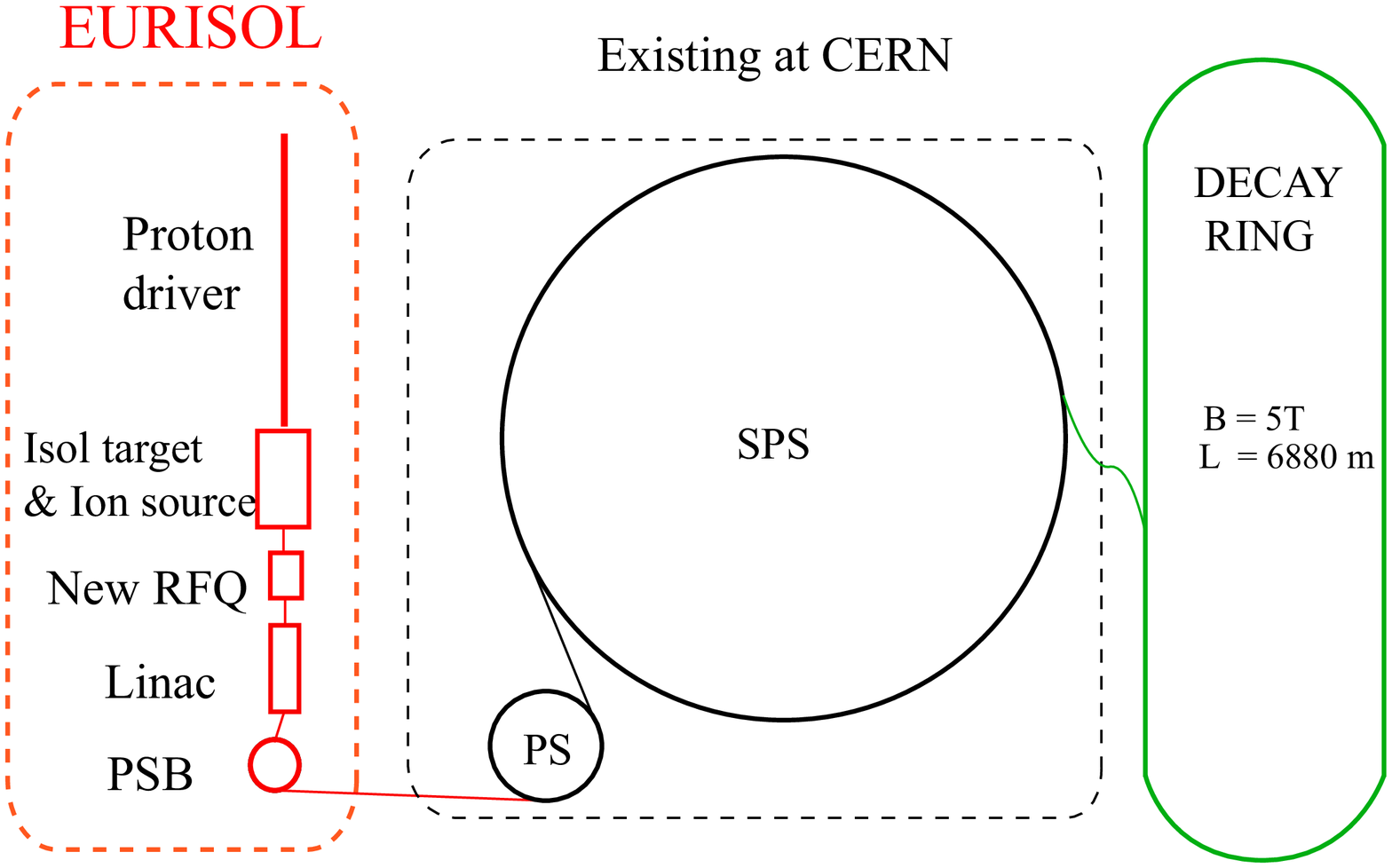,width=0.60\textwidth}  }
\mycaption{ A schematic layout of the BetaBeam complex. At left, the low energy part is
largely similar to the EURISOL project \cite{Eurisol}.
 The central part (PS and SPS) uses
existing facilities. At right, the decay ring has to be built.}
\label{fig:sketch}
\end{figure}

{\bf Ion production} 
Protons are delivered by a high power Linac. Beta Beam targets need
100 $\mu$A proton beam, at energies between 1 and 2 GeV.

In case the
 Super Proton Linac (SPL) \cite{SPL} would be used,
Beta Beams could be fired to the same detector together with
a neutrino SuperBeam \cite{SPL-Physics}.
 SPL 
is designed to deliver 2mA of 2.2 GeV (kinetic energy) protons, in such
a configuration Beta Beams would use 10\% of the total proton intensity,
leaving room to a very intense conventional neutrino beam.

 The $^6$He target consists 
either of a water cooled  tungsten core or of
a liquid lead core which works as a proton to neutron converter
surrounded by beryllium oxide \cite{Nolen}, aiming for 10$^{15}$
fissions per second. 
$^{18}$Ne can be produced by spallation reactions,
in this case protons will directly hit a magnesium 
oxide target.
The collection and ionization of the ions is performed using the ECR technique
 \cite{Sortais}.

This stage could be shared with nuclear physicists aiming
to a source of radioactive ions of the same intensity to what needed by a 
Beta Beam. A design study has been recently approved by E.U.: Eurisol \cite{Eurisol},
where both nuclear and neutrino physics issues will be studied.

{\bf Ion acceleration} The CERN PS and SPS can be used to accelerate the ions.
There is a well established experience at CERN about ion accelerators.
Ions are firstly accelerated to 
MeV/u by a Linac and to 300 MeV/u, in a single batch of 150 ns,
by  a rapid cycling synchrotron .
16 bunches (consisting of 2.5 10$^{12}$ ions each in the case of \He) are then
accumulated into the PS, and reduced to 8 bunches during their acceleration
to intermediate energies.
The SPS will finally accelerate the 8 bunches to the desired energy
 using a new 40 MHz RF system and the existing 200 MHz RF
system, before ejecting them in batches of eight 10 ns bunches 
into the decay ring.
The SPS could accelerate \He ions at a maximum $\gamma$ value of
$\gamma_{\He}=150$.

{\bf Decay ring}
The decay ring has  a total length of 6880 m
 and straight sections of 2500 m each (36\% useful length for
ion decays).
These dimensions are fixed by the  need 
to bend \He ions up to $\gamma=150$ using 5 T superconducting
magnets.
Due to the relativistic time dilatation, the ion lifetimes reach several
minutes, so that stacking the ions in the decay ring is mandatory to get enough
decays and hence high neutrino fluxes. The challenge is then to inject ions
in the decay ring and merge them with existing high density bunches.
As conventional techniques with fast elements are excluded, a new scheme 
(asymmetric merging) was specifically conceived \cite{mergexp}.

Summarizing, the main features of a neutrino beam based on the
BetaBeams concept are:

\begin{itemize}
\item the beam energy depends on the $\gamma$ factor. The ion
accelerator can be tuned to optimize the sensitivity of the
experiment;
\item the neutrino beam contains a single flavor with an energy
spectrum and intensity known a priori. Therefore, unlike conventional
neutrino beams, close detectors are not necessary to normalize the fluxes;
\item neutrino and anti-neutrino beams can be produced with a
comparable flux;
\item differently from SuperBeams, BetaBeams experiments search for $\nu_e
\rightarrow \nu_\mu$\  transitions, requiring a detector capable to
identify muons and separate them from electrons.
Moreover, since
the beam does not contain $\nu_\mu$\  or $\bar{ \nu}_\mu$\ in the initial
state, magnetized detectors are not needed. This is in contrast with
the neutrino factories (see Sec.\ref{sec:Nufact}) where the determination of the muon
sign is mandatory.
\end{itemize}

\noindent A baseline study for  a Beta Beam complex (Fig.~\ref{fig:sketch})
has been carried out at CERN \cite{Lindroos}.
The reference $\beta$B fluxes  are $5.8 {\cdot} 10^{18}$ \He\ useful
decays/year and $2.2{\cdot}10^{18}$ \Ne\  decays/year if a single ion specie
circulates in the decay ring.

The water \v{C}erenkov could be a suitable technology for a large detector
The physics potential has been initially computed in \cite{beta,Donini:2004hu}
 for $\gamma_{\He}=60$,
$\gamma_{\Ne}=100$ and
with a 440 kton detector at 130 km, Memphys, see also section 3.3.1.

\begin{figure}
    \centerline{\epsfig{file=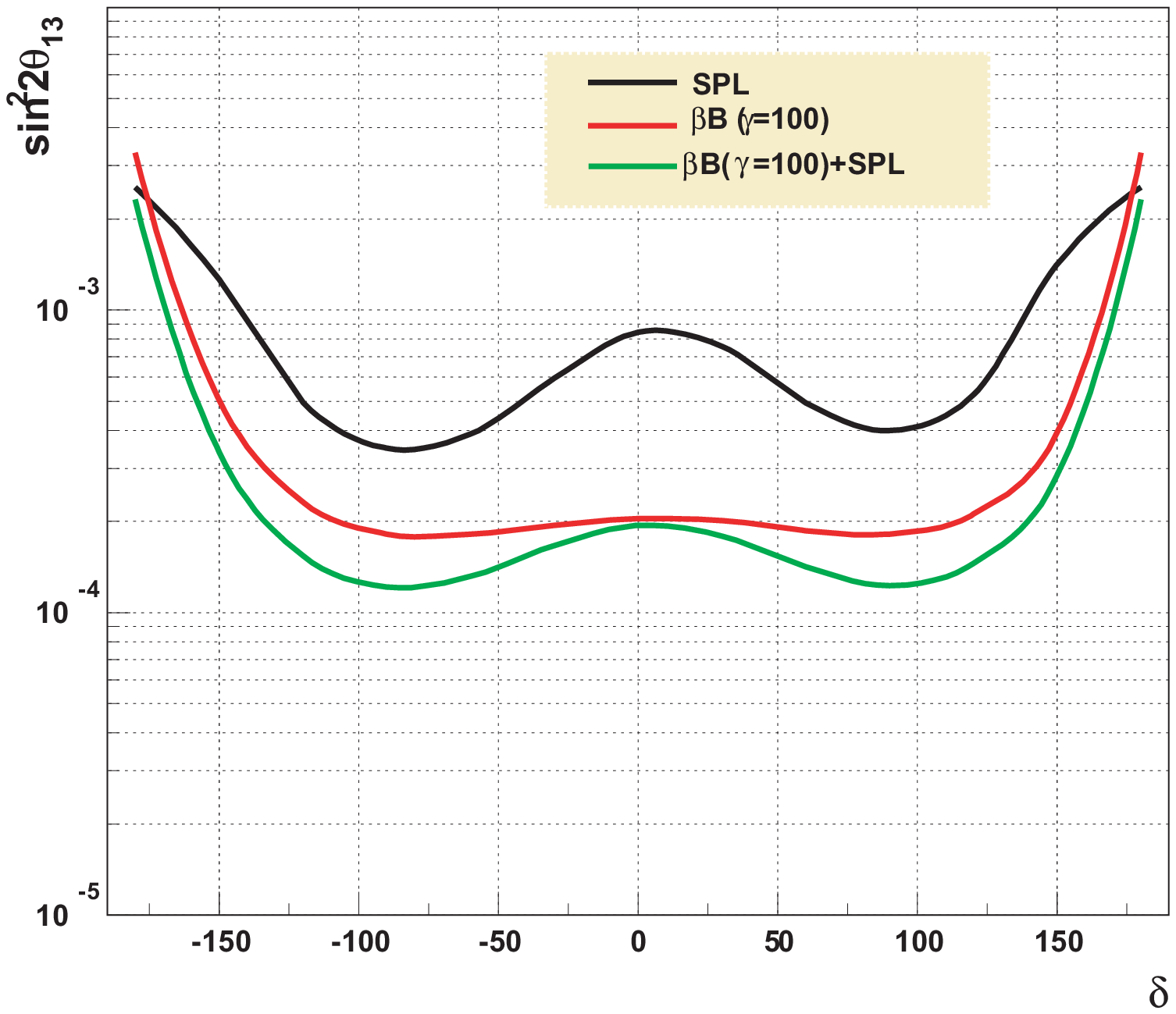,width=0.43\textwidth}}
    \mycaption{\thetaot \  
             sensitivity (90\% CL) as function of $\delCP$ for
             $\dmtt=2.5{\cdot}10^{-3}eV^2$, $\sigdm=1$, 2\%
             systematic errors.
             SPL-SB  sensitivities have been computed for a
             10 years \numu run, $\beta$B and $\beta$B$_{100,100}$
             for a 10 years \nue + \nubare\  run. 
      The SPL-SB  3.5 GeV, BetaBeam with $\gamma=100,100$ and their combination
       are shown.}
  \label{fig:th13}
\end{figure}

The most updated sensitivities for the baseline Beta Beam are computed in a
scheme where both ions are accelerated at $\gamma=100$, the optimal setup for
the CERN-Frejus baseline of 130 km,  \cite{MMNufact05}. 
   The \thetaot sensitivity curve,  computed with a 6 parameters fit minimized over the solar and the
   atmospheric parameters and projected
  over \thetaot, is shown in Fig.~\ref{fig:th13} \cite{MMNufact05}.
  Degeneracies induced by the unknown values of \sigdm and \thetatt
  are not accounted for in these first plots.

 The leptonic CP violation discovery potential (LCPV) has been computed with the
following procedure.
For any choice of a true value of \thetaot, $\bar{\theta}_{13}$, a loop on test
values of \delCP, $\bar{\delta}_{CP}$, is initiated, until the fit around
 ($\bar{\theta}_{13}, \bar{\delta}_{CP}$) is
  $3 \sigma$ away from any solution at $\delCP=0$ and
$\delCP=\pi$. While in ref.\cite{Donini:2004iv} this procedure is performed in the full
(\thetaot,\delCP) space (3 $\sigma$ corresponding to $\Delta \chi^2=11.8$),
 here the solution is searched having marginalized
out \thetaot (GLoBES function {\tt GblChiDelta} \cite{Globes}).
The LCPV at $3 \sigma$
  ($\Delta \chi^2= 9.0$) is shown in Fig.~\ref{fig:final}. It takes
 into account all the parameter errors and all
 the possible degeneracies \cite{MMNufact05}. As it is common practice in
 literature $\theta_{23}=40^\circ$
 has been used, to leave room for the octant ($\pi/2-\thetatt$) degeneracy.
 Each of the 4 true values of \sigdm and \thetatt:  
 normal: \sigdm=1, $\thetatt<\pi/4$;
 sign: \sigdm=-1, $\thetatt<\pi/4$; 
 octant: \sigdm=1, $\thetatt>\pi/4$;
 mixed: \sigdm=-1, $\thetatt>\pi/4$.
 has been fitted with the 4 possible fit combinations,
the worst case is then taken.
 Also shown are the leptonic CP violation discovery potentials
 neglecting the degenerate solutions (that is choosing the right combination
of \sigdm and \thetatt for the fit).
 Effect of degeneracies are sometimes visible for high values
 of \thetaot, precisely the region
 where they can be reduced by a combined analysis with atmospheric neutrinos
 \cite{Schwetz}. A quantitative computation of the combined analysis of Beta
Beam and atmospherics, as well as SPL superbeam and atmospherics, has been
recently shown in reference \cite{Schwetz2}. \\

BetaBeams require a proton driver in the energy range of 1-2 GeV, 0.5 MWatt power.
The SPL can be used as injector, at most 10\% of its protons would be consumed.
This allows a simultaneous $\beta$B and SPL-SB run, the two neutrino
beams having similar neutrino energies (cfr. Fig.~\ref{fig:fluxes}).
The same detector could then be
exposed to $2{\times} 2$ beams (\numu  and \nubarmu\  ${\times}$
\nue\  and \nubare) having access to CP, T and CPT violation searches in the same run.
This is particularly important because CP and T channels would have different 
systematics and different backgrounds, allowing for independent checks of the
same signal. Furthermore, the SPL $\nu_\mu$\ and \nubarmu\ beams would be the
ideal tool to measure signal cross sections in the close detector.

With this combination of neutrino beams a sensitivity to $\sin^2{2 \thetaot} \geq
2 \cdot 10^{-4} \; {(\rm 90\% CL)}$
exploiting a CP violation discovery potential
at 3 $\sigma$ if $\delCP \geq 18^\circ$  and $\theta_{13} \geq 0.55^\circ$ \cite{MMNufact04}

BetaBeam capabilities for the maximum values of $\gamma$ available with the SPS,
$\gamma\He=150$ have been computed in \cite{latestJJ}.

BetaBeam capabilities for ions accelerated at higher energies than
those allowed by SPS have been first computed in \cite{HighEnergy}
and subsequently in
\cite{latestJJ,HighEnergy2,HuberBB}.  These studies assume that
the same ion fluxes of the baseline scenario can be maintained.
However, this is not the case if the number of stored bunches is kept
constant in the storage ring. On the other hand, by increasing
$\gamma$ (i.e. the neutrino energy) the atmospheric neutrinos background
constraint on the total bunch length \cite{BetaBeam} becomes less
essential because of the reduced atmospheric neutrino flux at higher
energies.
Studies are in progress at CERN in order
to define realistic neutrino fluxes as a function of
$\gamma$\ \cite{MatsPrivate}.

\begin{figure}
\epsfig{file=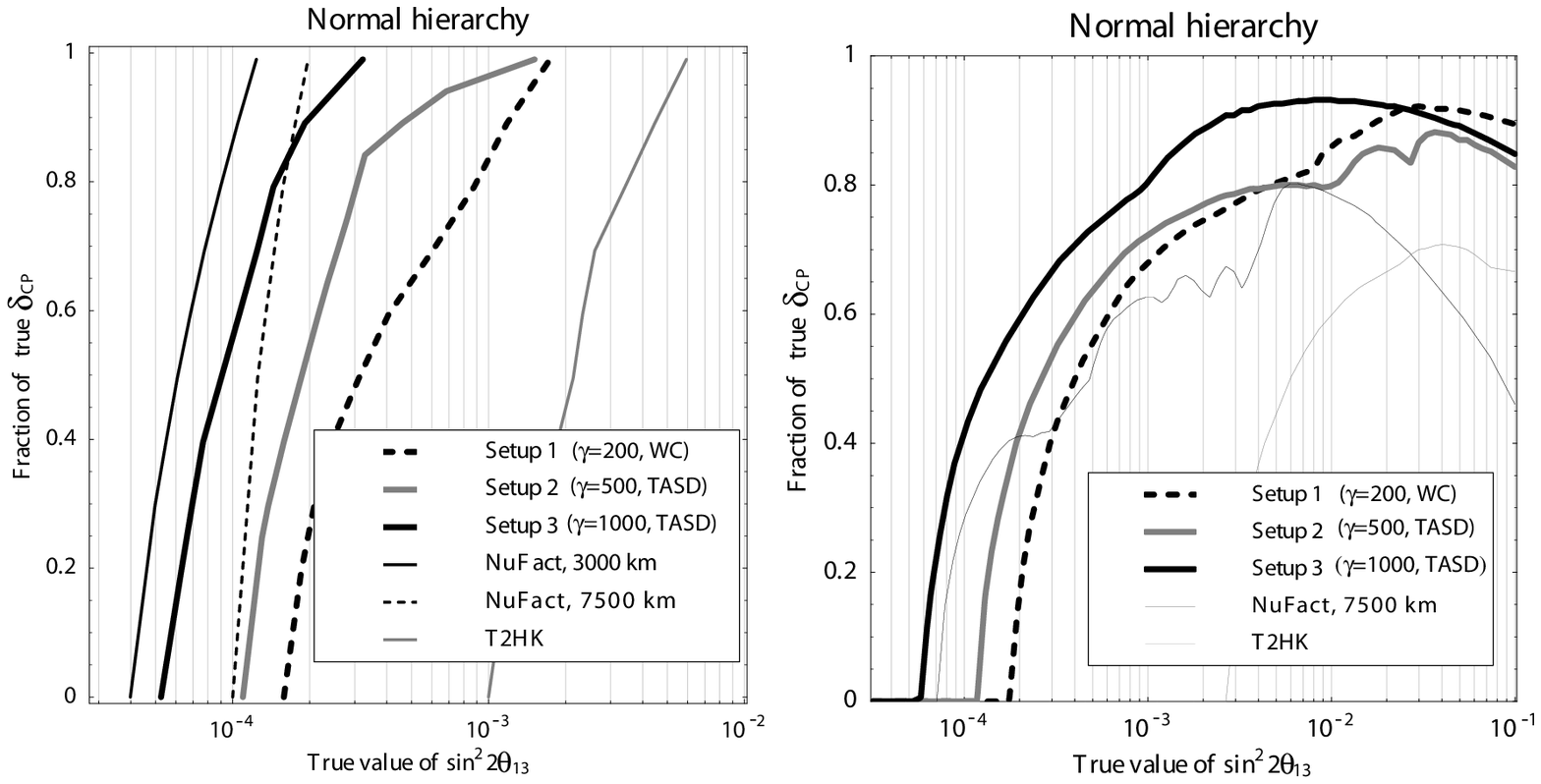,width=0.9\textwidth} 
\vspace*{-1.9cm}
\mycaption{\label{fig:t13cpf}LEFT: The $\stheta$ discovery reach (including
systematics and correlations) for different setups (the NO$\nu$A TASD detector is described
in Sections 2.2 and 3.1 ) as function of the 
true values of $\stheta$ and $\deltacp$ ($3 \sigma$ confidence level).
 The values of $\deltacp$ are ``stacked'' to the 
fraction of $\deltacp$, i.e., $\stheta$ will be discovered for a certain 
fraction of all possible values of $\deltacp$. For a uniform probability 
contribution in $\deltacp$, the fraction of $\deltacp$ directly 
corresponds to the probability to discover $\stheta$.
RIGHT: The sensitivity to CP violation for the normal
mass hierarchy for different experiments as function
of the true values of $\stheta$ and $\deltacp$ at the $3 \sigma$ confidence
level.  Both plots from reference \cite{HuberBB}}
\end{figure}

The outcome of these studies shows anyway that
higher energy Beta Beams have a CP discovery potential competitive with a
neutrino factory,
as shown in the plots of Fig.~\ref{fig:t13cpf}  \cite{HuberBB}.

 It is worth noting that if a high intensity Beta
Beam with $\gamma\sim 300\div500$ (requiring a higher energy
accelerator than SPS, like the Super-SPS\cite{SuperSPS}) can be built, a 40 kton
iron calorimeter located at the Gran Sasso Laboratory will have the
possibility to discover a non vanishing $\delta_{CP}$
if $\delCP>20^\circ$ for
$\theta_{13}\ge2^\circ$ (99\% C.L.) and measure the sign of $\Delta m^2_{23}$\ 
\cite{MigNufact05}. \\
 For a review on BetaBeams  see also
 \cite{BetaReview}.

A very recent development of the Beta Beam concept is the
conceptual possibility to have monochromatic, single flavor neutrino beams
thanks to the electron capture process \cite{Bernabeu,Sato}.
A suitable ion candidate exists: $^{150}$Dy, whose performances have
been already delineated \cite{Bernabeu}.

\subsection{
\label{sec:Nufact}The Neutrino Factory}

\begin{figure}
  \begin{center}
  \mbox{\epsfig{file=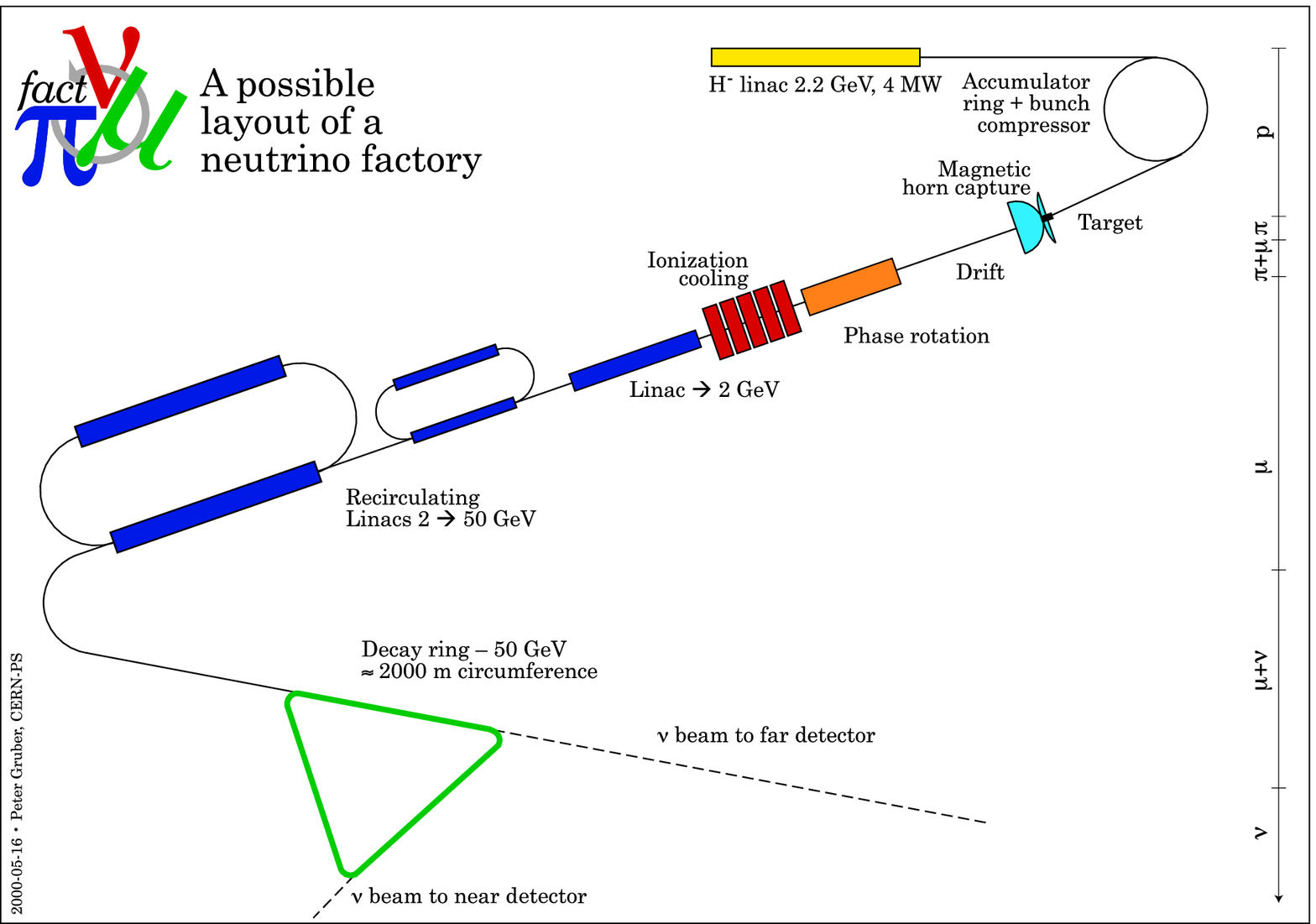,width=0.7\textwidth}}
  \end{center}
  \vskip -0.3cm
  \mycaption{Schematic layout of a Neutrino Factory.}
  \label{fig:nufact}
 \end{figure}

\footnote{Material for this Section is mainly taken from ref.~\cite{Blondel05}}
In a Neutrino Factory \cite{nufact} muons are accelerated from an intense source to
 energies of several GeV, and injected in a storage ring with long straight sections.
 The muon decays:
$$
  \mu^+ \rightarrow e^+\nu_e\nubarmu \quad {\rm and} \quad \mu^- \rightarrow e^-\nubare 
\nu_\mu
$$
provide a very well known flux with energies up to the muon energy itself.
 The overall layout is shown in Fig.~\ref{fig:nufact}. 

Neutrino Factory designs have been proposed in Europe \cite{Aut99,Gru02},
 the US \cite{MuColl,StudyI,StudyII}, and Japan \cite{Japnufact}.
 Of these designs,
 the American one is the most developed, and we will use it as an example
 in general with a few exceptions. The conclusions of these studies is that,
 provided sufficient resources, an accelerator complex 
capable of providing about $10^{21}$ muon decays per year can be built. 
The Neutrino Factory consists of the following subsystems: 

{\bf Proton Driver.}  Provides 1-4 MW of protons on a pion production target.
 For the Neutrino Factory application the energy of the beam within 4-30 GeV 
is not critical,
 since it has been shown that the production of pions is roughly proportional
 to beam power. The time structure of the proton beam has to be matched with
 the time spread induced by pion decay (1-2 ns); 
for a Linac driver such as the SPL,
 this requires an additional accumulator and compressor ring. 

{\bf Target, Capture and Decay.} A high-power target sits within a 20T
 superconducting solenoid, which captures the pions.
 The high magnetic field smoothly decreases to 1.75T downstream of the target,
 matching into a long solenoid decay channel. A design with horn collection 
has been proposed at CERN for the Neutrino Factory, with the benefit that it
 can be also used for a SuperBeam design. 

{\bf Bunching and Phase Rotation.} The muons from the decaying pions are 
bunched using a system of RF cavities 
with frequencies that vary along the channel.
 A second series of RF cavities with higher gradients is used to rotate the
 beam in longitudinal phase-space, reducing the energy spread of the muons.

{\bf Cooling}.  A solenoid focusing channel with high-gradient 201 MHz RF cavities
 and either liquid-hydrogen or LiH absorbers is used to reduce the transverse
 phase-space occupied by the beam. The muons lose, by dE/dx  losses,  
both longitudinal- and transverse-momentum as they pass through the absorbers.
 The longitudinal momentum is restored by re-acceleration in the RF cavities.

{\bf Acceleration.} The central momentum of the muons exiting the cooling channel 
is 220 MeV/c. A superconducting Linac with solenoid focusing is used to raise
 the energy to 1.5 GeV. Thereafter, a Recirculating Linear Accelerator raises
 the energy to 5 GeV, and a pair of Fixed-Field Alternating Gradient rings
 accelerate the beam to at least 20 GeV.

{\bf Storage Ring.} A compact racetrack geometry ring is used, in which 35\% of the
 muons decay in the neutrino beam-forming straight section. If both signs are accelerated, one can inject in two superimposed rings or in two parallel straight sections. 
This scheme produces over $6 \cdot 10^{20}$ useful muon decays per operational year and
 per straight section in a triangular geometry.\\

 The European Neutrino Factory design
 is similar to the US design, but differs in the technologies chosen to
 implement  the subsystems.

 The Japanese design is quite different, and uses very large acceptance 
accelerators. Cooling, although it would improve 
performance, is not considered mandatory in this scheme.  

An important Neutrino Factory R\&D effort is ongoing in Europe, Japan, and the
 U.S. since a few years. Significant progress has been made towards optimizing
 the design, developing and testing the required components, and reducing the cost.

 To illustrate this progress, the cost estimate for a recent update of the
 US design \cite{APS04} is compared in Table~\ref{tab:nufcosts}  with the corresponding cost for 
the previous "Study II" US design \cite{StudyII}.
In this design the Neutrino Factory would accelerate protons up to 20 GeV/c, with
a flux of $1.2 \cdot 10^{20}$ muon decays per straight section per year for a proton driver
power of 1 MW ($4.8 \cdot 10^{20}\;\mu$ decays year at 4 MW) . 

 It should be noted that the
 Study II design cost was based on a significant amount of engineering
 input to ensure design feasibility and establish a good cost basis. 
Neutrino Factory R\&D has reached a critical stage in which support is
 required for two key international experiments (MICE \cite{MICE} and 
Targetry \cite{target-exp}) and a third-generation international design study.
 If this support is forthcoming, a Neutrino Factory could be added to
 the Neutrino Physics roadmap by the end of the decade. 

\begin{table}
  \mycaption{Comparison of unloaded Neutrino Factory costs estimates in M\$ 
   for the US Study II design and improvement estimated for the latest
   updated US design (20 GeV/c muons). Costs are shown including A: the whole complex;
   B no Proton Driver and C:  no proton driver and no Target station in the estimates.
    Table  from Ref.~\cite{APS04}.}
  \begin{center}
  \begin{tabular}{|l|c|c|c|}
    \hline
    Costs in M\$ & A & B  & C \\
    \hline
    Old estimate from Study II & 1832 & 1641 & 1538 \\
    \hline
    Multiplicative factor for new estimate & 0.67 & 0.63 & 0.60 \\
    \hline
  \end{tabular}
  \end{center}
  \label{tab:nufcosts}
\end{table}

\subsubsection{Oscillations physics at the Neutrino Factory}  
Considering a Neutrino Factory with simultaneous beams of positive and negative muons,
  12 oscillation processes can in principle be studied, Table~\ref{tab:procs}.   

\begin{table}[h!]
\begin{center}
\mycaption{Oscillation processes in a Neutrino Factory}
\begin{tabular}{|c|c|c|}
\hline
  $\mu^+ \rightarrow e^+\nu_e\nubarmu$ &  $\mu^- \rightarrow e^-\nubare$  & \\
\hline
$\nubarmu \rightarrow \nubarmu$  & $\nu_\mu \rightarrow \nu_\mu$   & disappearance \\
$\nubarmu \rightarrow \nubare$   & $\nu_\mu \rightarrow \nu_e$     & appearance (challenging) \\
$\nubarmu \rightarrow \nubartau$ & $\nu_\mu \rightarrow \nu_\tau$  & appearance (atm. oscillation) \\
$\nu_e \rightarrow \nu_e$        & $\nubare \rightarrow \nubare$   & disappearance \\   
$\nu_e \rightarrow \nu_\mu$      & $\nubare \rightarrow \nubarmu$  & appearance: ``golden'' channel \\  
$\nu_e \rightarrow \nu_\tau$     & $\nubare \rightarrow \nubartau$ &  appearance: ``silver'' channel \\ 
\hline
\end{tabular}
\label{tab:procs}
\end{center}
\end{table}

 Of course the neutrinos coming from decays of muons of different charge
 must no be confused with each other, this can be done by timing provided
 the storage ring is adequately designed. 

One of the most striking features of the Neutrino Factory is the precision 
with which the characteristics of all components of the beam could be known.
This was studied extensively in a CERN 
Report~\cite{ECFA-report}, where the following effects were considered
\begin{itemize}
\item
beam polarization effects, and its measurement by a polarimeter, allowing 
extraction of the beam energy, energy spread and verification that the 
polarization effects on the neutrino fluxes average out to zero 
with high precision;
\item
beam divergence effects, with the preliminary, conceptual study of a 
\v{C}erenkov device to monitor the angular distribution of muons in the beam
\cite{Broncano}
\item
radiative effects in muon decay;
\item
absolute normalization to be obtained both from a beam monitor, 
with the added possibility of an absolute cross-section normalization 
using the inverse muon decay reaction, 
$\nu_{\mu} e^- \rightarrow \mu^- \nu_e$, in the near detector;  
\end{itemize}
with the conclusion that, in principle, a normalization of fluxes and 
cross-sections with a precision of $10^{-3}$ can be contemplated. 
Some of these features should also be present for a BetaBeam, 
and for any facility in which a stored beam of well defined optical properties
is used to produce neutrinos. This is an essential difference with respect
to the SuperBeams, where the knowledge of relative neutrino-vs-antineutrino
cross-sections and fluxes will rely on
the understanding of the initial particle production.  

 The Neutrino Factory lends itself naturally 
to the exploration of neutrino oscillations
between $\nu$ flavors with high sensitivity. 
The detector should be able to perform both
appearance and disappearance experiments, providing lepton
identification and charge discrimination which is a tag for the initial
flavor and of the oscillation.
   In particular the search  for $\nu_e \rightarrow  \nu_{\mu}$ transitions (``golden channel") \cite{Golden}
  appears to be very attractive at the Neutrino Factory,  because
  this transition can be studied in appearance  mode looking for
  $\mu^-$ (appearance of wrong-sign $\mu$)
  in  neutrino beams where the neutrino type that is searched for
  is totally absent ($\mu^+$ beam in {\large $\nu$F}).
  
The emphasis has been placed so far on small mixing angles and
small mass differences. With two 40 kton magnetic detectors (MINOS like) 
at 700 (or 7000) and 3000 km,  
with a conservative high energy muon detection threshold of 5 GeV, 
exposed to both polarity
  beams and $10^{21}$ muon decays, it will be possible to explore
  the $\theta_{13}$ angle down to $0.1^\circ$ opening the possibility
  to measure the $\delCP$ phase 
  \cite{Golden,PilarNufact, Huber03}, as it is shown by the plots of 
 Fig.\ref{fig:final}.

On the other hand,  the relative high energies of neutrinos selected 
by placing such a high threshold on muon energies 
require very long
  baselines (several thousands kilometers) for Neutrino Factories experiments,
   and at these baselines
  CP asymmetries are dominated by matter effects \cite{Geer}. 
     Taking advantage of the matter effects, such an experiment will determine
      unambiguously sign($\Delta m^2_{23}$) for large enough $\theta_{13}$ 
     ($\theta_{13} \geq 2^\circ$). However, we remind that, such as for other
     facilities, the determination of ($\theta_{13},\delta$) at the Neutrino 
     Factory is not free of ambiguities: up to eight different regions of 
     the parameter space can fit the same experimental data. 
     In order to solve  these ambiguities, a single experimental point 
on a single neutrino beam is not enough. 

One possibility at the Neutrino Factory is to make use of the 
rich flavour content of the beam. This imply an optimized
network of detectors with different characteristics. 
Indeed, a specific disadvantage of the considered magnetized
iron detector when dealing with degeneracies is the following: 
the lower part of the Neutrino Factory spectrum 
(say, $E_\nu \in [0,10]$ GeV) cannot be used due to the
extremely low efficiency in this region of the detector.
     This part of the spectrum, on the other hand, is 
     extremely useful to solve degeneracies, as it has been shown in 
     several papers \cite{Huber02}.


One possibility is to envisage that all of  BetaBeams,
SuperBeams and Neutrino Factories will be available. 
Several investigations on how to solve this
problem have been carried out, as reported in~\cite{Donini:2003kr} and
references therein. 

A more interesting but challenging task will be to assume that 
only one of these facilities will become available 
(a more economical assumption!) and to 
investigate its ability to solve these ambiguities. 

\begin{figure}
   \begin{center}
     \epsfig{file=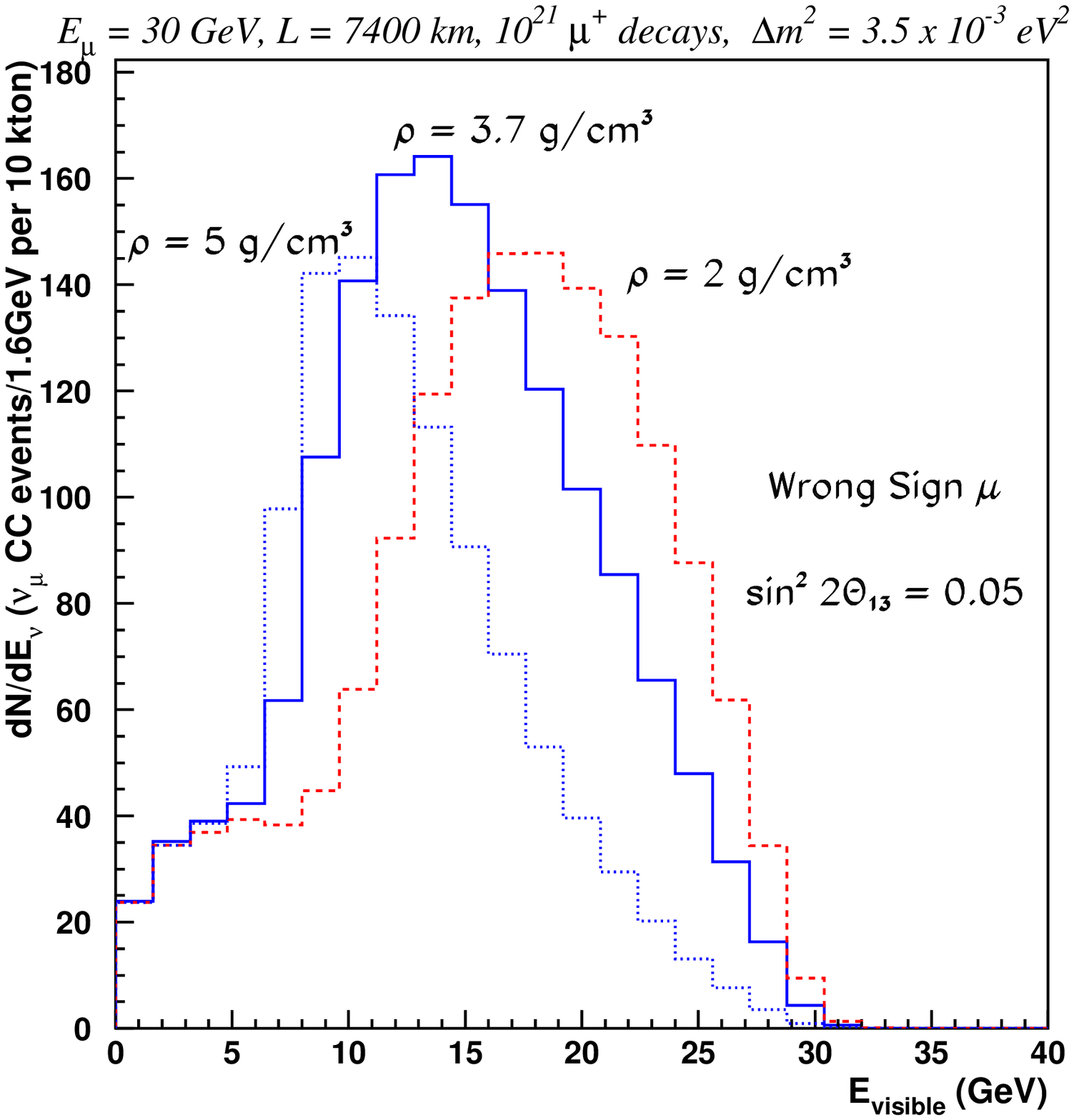,width=0.45\textwidth} \hfill
   \end{center}
  \vspace*{-0.7cm}
    \mycaption{Variation of the MSW resonance
   peak for wrong sign muons as a
   function of Earth density. The plot is
   normalized to $10^{21} \mu^+$ decays.
   From reference\cite{Bueno00}.
   }
    \label{matter-res}
\end{figure}

There are several handles to this problem at a Neutrino Factory. 
Clearly one should use more than just the wrong sign muons. 
Such a study was performed assuming the feasibility of a liquid argon detector
\cite{Bueno00}. By separating the events into several classes, 
right sign muon, wrong sign muon, electron and neutral current; 
and by performing a fine energy binning down to low energies; 
it was shown that the matter resonance could be neatly measured as shown in 
Fig.~\ref{matter-res}. The simultaneous observation of the four 
aforementioned channels 
was shown to allow resolution of ambiguities to a large extent.

\begin{figure}
\begin{center}
  {\epsfig{file=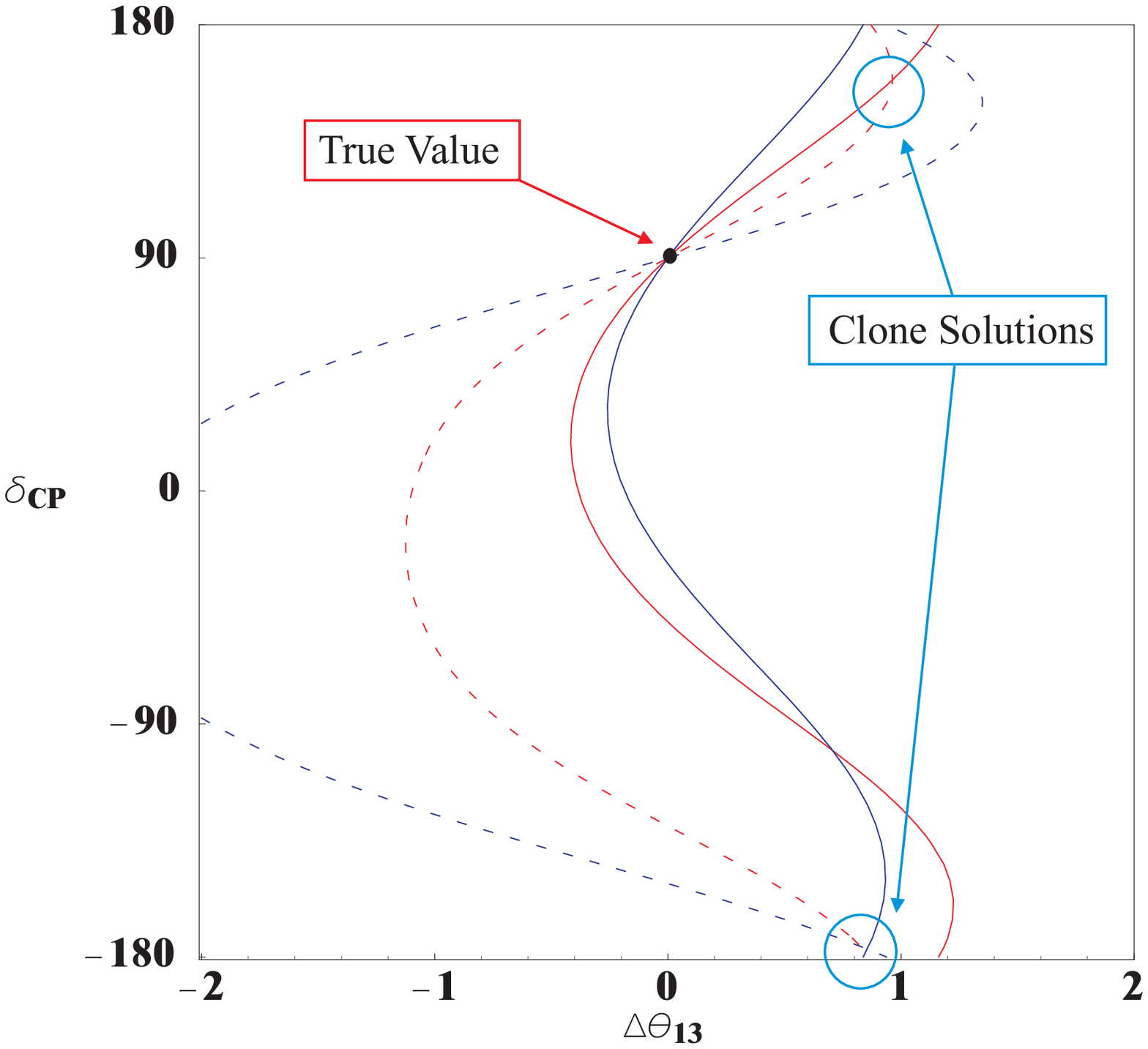,width=0.45\textwidth}} \hfill
  {\epsfig{file=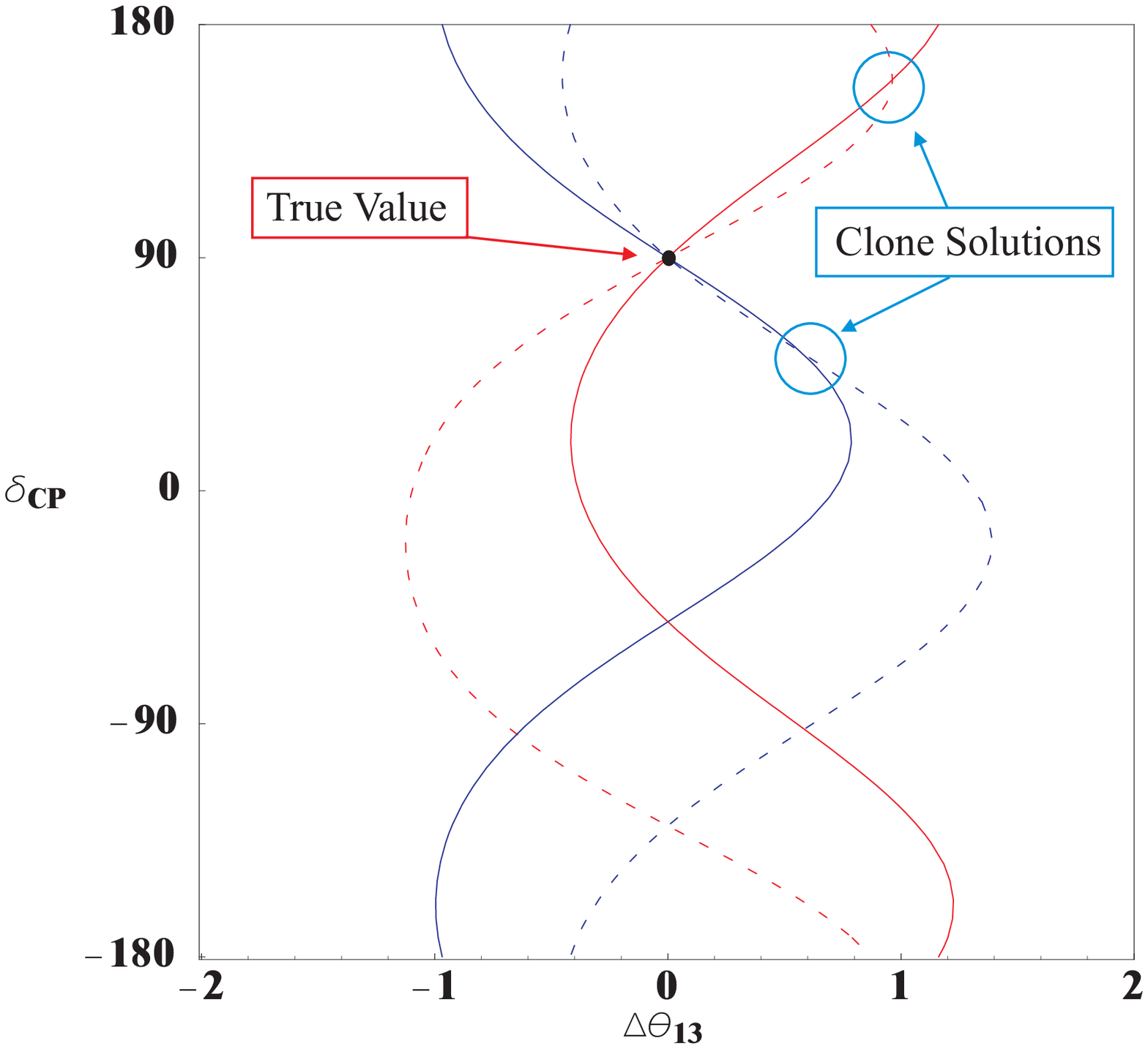,width=0.45\textwidth}}
\end{center}
  \vspace*{-0.4cm}
    \mycaption{Solving the intrinsic degeneracy: 
two baseline L=730 and 3500 km, same channel example on the left,
vs two channels $\nu_e \rightarrow \nu_{\mu}$ vs $\nu_e \rightarrow \nu_{\tau}$
same baseline example on the right. From Ref.~\cite{silverRigo}
}
    \label{degeneracies}
\end{figure}

The tau appearance channel {\em{silver channel}}~\cite{silver} 
has been advocated 
as a powerful means of solving ambiguities. This can be readily 
understood since this channel
has the opposite-sign dependence on ${\delta_{CP}}$ than the golden one, 
while having similar dependence on matter effects and ${\theta_{13}}$. 
Another channel to be used is the $\nu_\mu$ disappearance channel,
rather effective for large values of \thetaot \cite{Donini:2005db}.
The principle of degeneracies-solving using  several baselines, 
binning in energies and both silver and golden channels 
is explained on Fig.~\ref{degeneracies}. The full demonstration
 that a Neutrino Factory alone with a complete set 
of appropriate detectors and two baselines could unambiguously 
do the job remains however to be worked out. 

According to Table~\ref{tab:procs}, Neutrino Factory potential could
be further improved with a detector capable of measuring the charge of the
electrons. R\&D efforts for a liquid argon detector embedded in a magnetic
fields are ongoing \cite{RubbiaNufact}; the first curved tracks were recently 
observed in a 10 liters Liquid Argon TPC embedded in magnetic 
field~\cite{Badertscher05}.

\section{Research and development on detectors: status and priorities}

\subsection{Low-Z Tracking Calorimetry}

Low-Z Tracking Calorimetry optimizes the detection of
electrons in the final state by using a fine sampling in terms of radiation lengths, leading to the choice for a low-Z passive material. This is the technique used for the study of $\nu_{\mu}
e^-$ scattering in the CHARM II experiment at CERN \cite{3}. Here we discuss the status of the design of the NO$\nu$A experiment proposed for the detection of  $\nu_{\mu} - \nu_e$ oscillations in the off-axis NuMI
beam at Fermilab \cite{Nova}, with the observation of $\theta_{13}$ as its prime aim. 

The No$\nu$A detector foresees a mass one order of magnitude larger than MINOS
\cite{Minos} and at the same time, in relation to its aim, a finer sampling ($\Delta X_0 < 0.3$ if particle board is used as passive material,
to be compared to 1.5 with MINOS, which has iron as the passive material).
 The principal technical issue is to improve the performance and substantially reduce the unitary cost of the trackers. 
The main technological innovation consists in the use of liquid scintillator
 read by Avalanche Photodiodes (APDs), instead of plastic scintillator read
 by multianode PMTs. 

 The detector finally chosen for NO$\nu$A is a "Totally Active" Scintillator Detector (TASD),
 with a 30 kton mass, of which 24 kton are of liquid
 scintillator and the rest of PVC. In TASD, the scintillator modules have 
 cells along the beam 6.0 cm long  and  3.9 cm width.
 TASD consists of a single block with overall dimensions 15.7x15.7x132 $m^3$.
 Lacking the particle board, the PVC must provide 
a self-supporting structure for a detector as high as a five-storey building.
 Since last year, progress has been made also in the mechanical design and
 in the assembling methods.  

 The electromagnetic energy resolution is $\Delta E/E \sim
10\%/\sqrt{E(GeV)}$, the
 almost continuous pulse height information along the track helps in $e/\pi^0$
 discrimination.

The use of APDs results in a considerably lower cost
than with PMTs. 2x16 pixel APDs are commercially produced in large quantities and already foreseen for the CMS experiment at LHC. The high
quantum efficiency, about 85\%, allows to have longer strips and less readout channels.  

The No$\nu$A design is in constant progress.
 If funding would begin in late 2006, the No$\nu$A detector could be ready in
 2011.
 It is worth noting that the progress with the development of the trackers is
 potentially useful also for magnetized iron spectrometers for neutrino
 factories or colliders.

\subsection{Magnetized Iron Spectrometers}

This technique is conventional, but the mass to be considered
is one order of magnitude larger than for present magnetized iron spectrometers, like MINOS. 

Recent studies indicates that a magnetized iron toroidal spectrometer of the required mass is feasible \cite{Nelson}.
 On one hand, the design of toroids with radius up to 10 m can be extrapolated from MINOS, with thicker plates for larger planes. On the other hand, the  No$\nu$A liquid scintillator technology with APD readout allows to have transverse dimensions twice as large than in MINOS. Such a detector concept permits direct use of the experience with MINOS.

The India Neutrino Observatory (INO) \cite{India}
 foresees a dipole magnet equipped with RPCs, as in MONOLITH \cite{17}.
 The INO basic motivation is the study of atmospheric neutrinos like with
 MONOLITH; a future use in a very long baseline experiment with a
 $\nu$-factory is envisaged. The investigations started with MONOLITH on the
 detector performance for design optimization have to be continued,
 in a comparison with other detectors. 

A conceptual spectrometer  based on a 40 kton iron solenoid magnetized at 1 T by a superconducting coil, with embedded solid scintillator rods as the active detectors, has been presented in Ref. \cite{16}. Considerable more work is required to define its features and assess its practical feasibility.

In general, practical problems (mechanics, magnet design, etc.) must be thoroughly addressed. In addition, as already mentioned more simulation work is required in order to understand and optimize the performance by a proper choice of the main detector parameters.   

\subsection{Water $\mathrm{\check{C}}$erenkov}

Water $\mathrm{\check{C}}$erenkovs can provide a very large target
mass and, if the photo-sensors have a
sufficient density, a sensitivity down to the low energies
of solar neutrinos. Its capabilities concern $\nu$ astrophysics, $\nu$ oscillations and proton decay. Above a few GeV, DIS dominates over QE scattering and leads to frequent multi-ring events more complicated to reconstruct. A similar limitation in energy comes from difficulties in the $e/\pi^0$ discrimination at high energies. The technique is not suitable at the high energies of $\nu$-factories, where, in addition, a muon charge measurement is needed. One should also remark that the low neutrino cross-section at low energies reduces the advantage given by the very large mass which can be realized.

The detectors presently under study represent the third generation of
successful detectors, with in each stage an increase by one order of magnitude in mass. The performance of Super-KamiokaNDE has been widely simulated and
observed, providing a basis for a mass extrapolation by one order of magnitude. The performance as well as the limitations are well known, also from K2K and related tests. 

Two detector designs are being carried out, namely Hyper-KamiokaNDE
\cite{7} and UNO \cite{8}. The design of a detector to be
located at Frejus (Memphys) has been also initiated.

Hyper-KamiokaNDE foresees two 500
kton modules placed sideways, each consisting of five 50 m long optical compartments.
The cost is higher than for a single
module, but maintenance with one module always alive and a staging 
in the detector construction become possible. The present design foresees about 200,000 20" PMTs, to be compared with 11,146 in Super-KamiokaNDE. The detector could be constructed in about 10 years, starting after a few years of T2K operation. 

The UNO design provides a 650 kton mass subdivided into three optical compartments with
 different photo-sensor coverage. The central one has a 40\% coverage
 as in Super-KamiokaNDE, allowing to pursue solar $\nu$ studies. The side compartments have 10\% coverage. The number of 20" PMTs is two times smaller than with 40\% coverage for the full detector, but still amounts to 56,650. The question arises as to whether this subdivision, with its non-uniformity given by the lower coverage in $\frac{2}{3}$ of the detector, is the optimal solution to reduce the global cost.
 
By giving appropriate aspect ratio and shape to the cavern, its realization does not seem a problem.

A large fraction of the total detector cost, reaching $\sim \frac{1}{2}$ 
or more if PMTs are used, comes from the photo-sensors. The present cost of 100,000-200,000 20" PMTs is hard to cover. Moreover, their production would take about 8 years, leading also to storage problems. The main issue is thus the development and acquisition of photo-sensors at acceptable cost and production rate, as well as an improvement of their performance. A better time resolution would improve neutrino vertexing and single photon sensitivity would give better ring reconstruction. A strong collaboration with industry is essential, as for the development of 20" PMTs by Hamamatsu for KamiokaNDE and Super-KamiokaNDE.

The Hamamatsu 20" glass bulbs are manually blown by specially trained people. Automatic
manufacturing does not seem a practical solution to reduce the cost and speed-up the production rate, as the required quantity is still small compared to commercial PMTs. The question is whether a size smaller than 20", with an appropriate coverage, is more practical and cost effective. With a smaller size, automatic bulb manufacturing is eased and the risk of implosion decreased, with a possible saving in the plastic protection to damp implosions. For R\&D on PMTs, collaborations have been established with industries also in Europe (Photonis) and USA.

To explore alternatives to PMTs, studies on new photo-sensors have been launched. In addition to reduce cost, while improving production rate and performance, it is essential to achieve the long term stability and reliability which is proven for PMTs. 

Hybrid Photo-Detectors (HPD) are being developed by Hamamatsu, in collaboration with ICRR of Tokyo University. The HPD glass envelope is internally
coated with a photo-cathode and a light reflector. Electrons are
accelerated by a very high voltage towards an Avalanche Diode (AD). The strong
electron bombardment results in a high gain ($\sim$ 4500 for 20 kV voltage)
in this first stage of amplification. It gives a remarkable single photon
sensitivity and makes ineffective the AD thermally generated noise. The gain is lower than with PMTs, hence stable and
highly reliable amplifiers are needed. The degradation in time resolution given by the transit time spread through the dynode chain is avoided, so that a $\sim$ 1 ns time resolution can be achieved, to be compared with the 2.3 ns of the 20" PMTs. The cost reduction with
respect to PMTs essentially comes from the use of solid state
devices like the AD, avoiding the complicated PMT dynode structure.

The principle has been proved with a 5" HPD prototype. Successful results from tests of an 13"
prototype operated with 12 kV are now available, showing a $3 \cdot 10^4$ gain, good single photon sensitivity, 0.8 ns time resolution and a satisfactory gain and timing uniformity over the photo-cathode area. The next step will be the operation at a voltage up to 20 kV, giving a higher gain and a wider effective area of the photo-cathode. The development of HPD has been initiated also in Europe, in collaboration with Photonis.

\subsubsection{The MEMPHYS detector}
\label{sec:Memphys}
The MEMPHYS (MEgaton Mass Physics) detector is a Megaton-class water
 \v{C}erenkov in the straight extrapolation of the well known and robust technique
 used for the Super-Kamiokande detector. It is designed to be
 located at Frejus, 130~km from CERN
 and it is an alternative design of the UNO \cite{UNO} and Hyper-Kamiokande \cite{T2K}
 detectors and shares the same physics case both from non accelerator domain
 (nucleon decay, SuperNovae neutrino from burst event or from relic explosion, 
solar and atmospheric neutrinos) and from accelerator (SuperBeam, BetaBeam) domain.
 For the physics part not covered by this document, this kind of megaton water
 detector can push the nucleon decay search up to $10^{35}$~yrs in the $e^+\pi^0$
 channel  and up to few $10^{34}$~yrs in the $K^+\overline{\nu}$ channel,
 just to cite these benchmark channels. MEMPHYS can register as many as 150,000
 events from a SN at 10~kpc from our galaxy and 50~events or so from Andromeda.
 To detect relic neutrinos from past SuperNovae explosions one can use pure water
 and get a flux of 250~evts/10~y/500 kton or increase this number by a factor
 10 by adding gadolinium salt.

A recent civil engineering pre-study to envisage the possibly of large cavity excavation
 located under the Frejus mountain (4800~m.e.w.) near the present
 Modane Underground Laboratory has been undertaken.
 The main result of this pre-study is that MEMPHYS may be built with present
 techniques as 3 or 4 shafts modular detector, $250000~\mathrm{m}^3$
 each with 65~m in diameter, 65~m in height for the total water containment. 
Each of these shafts corresponds to about $5$ times the present Super-Kamiokande cavity.
 For the present physical study, the fiducial volume of 440~kton which
 means 3 shafts and an Inner Detector (ID) of 57~m in diameter and 57~m in height is assumed.
 Each ID may be equipped with photodetectors (PMT, HPD,...)
 with a surface coverage at least 30\%. 
The Frejus site, 4800~m.w.e, offers a natural protection against cosmic rays by a factor $10^6$.

The decision for cavity digging is fixed at 2010 after an intense Detector Design Study
 (eg. cavity excavation, photodetector R\&D)
 performed in parallel of the digging of at least a Safety Gallery in the Frejus
 road tunnel. One may note that this key date may also be decisive for SPL 
construction as well as for the choice of the EURISOL site. After that, the excavation
 and PMT production are envisaged to take seven years or so, and the non accelerator
 program can start before the rise up of the accelerator program
 (SuperBeam and BetaBeam) which may start before 2020.     

  A first estimate of the costs of such a
 detector is reported in Table~\ref{Memphys}
 \begin{table}[h]
 \mycaption{Preliminary cost estimate of the MEMPHYS detector}
 \begin{center}
 \begin{tabular}{|l|c|}
 \hline
   3 Shafts   &  240 ME \\
 \hline
  Total cost of 250k 12'' PMTs & 250 ME \\
 \hline
   Infrastructure & 100 ME \\
 \hline
   Total  & 590 ME \\
 \hline
\end{tabular}
\end{center}
\label{Memphys}
\end{table}

\subsection{Liquid Argon Time Projection Chamber}

The Liquid Argon Time Projection Chamber (LAr TPC) provides an excellent imaging device and a dense neutrino target. It is a true "electronic bubble chamber" with a much larger mass (3 ton for Gargamelle, up to 100 kton now envisaged for LAr TPCs).
The "state of the art" is given by the 300 ton T300 ICARUS \cite{10} prototype module, tested at ground level in Pavia but not yet used in an experiment.

Two mass scales are foreseen for future experiments. For close $\nu$
detectors in Super-Beams, a mass of the order of 100 ton is
envisaged. Detectors with 50-100 kton masses are proposed for $\nu$ oscillation,
$\nu$ astrophysics and proton decay \cite{11}\cite{12}\cite{13}, implying a step in mass by more than two orders of magnitude with respect to the
T300 ICARUS module.

Cryogenic insulation imposes a minimal surface/volume ratio. The modular ICARUS approach has thus to be abandoned for a single very large cryogenic module with about 1:1 aspect ratio.

To limit the number of readout channels, drift lengths have to be
longer than the 1.5 m of the T300 module. The LAr TPC envisaged for the off-axis NuMI beam \cite{11} \cite{12} foresees 3 m drift lengths
with readout conceptually as in ICARUS. In another approach \cite{13}, the
very strong attenuation over a much longer drift length (20 m) is compensated  
by the so-called "Double Phase" amplification and readout
\cite{14}, tested on the ICARUS 50 liter
chamber \cite{15}. In both cases, the signal attenuation imposes a
liquid Argon contamination by electronegative
elements at the 0.1 ppb level.

In Ref. \cite{13} a 100 kton detector consisting of a single module both for
cryogenics and readout has been proposed. The 20 m drift in a field raised to 1 kV/cm results in a 10 ms drift time. With a 2 ms electron live-time in liquid Argon, the
6000 electrons/mm signal is attenuated by $e^{-t/\tau}$
$\approx$ 1/150 and becomes too low for a readout as in ICARUS. In the
Double Phase readout, electrons are extracted from the
liquid by a grid and multiplied in the gas phase
using gas chamber techniques. The $\sigma \approx$ 3 mm
spread from the diffusion in the 20 m drift gives an intrinsic limit to the readout granularity.

The design of the single large cryostat \cite{13} can
benefit from the techniques developed for transportation and storage of
large quantities of liquefied  natural gas, kept at boiling temperature by letting it evaporate. A large
cryogenic plant is needed for the initial filling and for the continuous refilling to compensate the evaporation.

The R\&D for the detector
of Ref. \cite{13} foresees: drift under 3 atm pressure as at 
the tank bottom; charge extraction and amplification;
imaging devices; cryogenics and cryostat design, in collaboration with industry; a column-like prototype with 5 m drift and
double-phase readout, with a 20 m drift simulated by a reduced E
field and by a lower liquid Argon purity; test of a prototype in a
magnetic field (as for use in a $\nu$-Factory); 
underground safety issues. 

Tests on a 10 liter LAr TPC inserted in a 0.55 T magnetic field have been performed \cite{Andre10l}, together with a tentative layout for the implementation of a large superconducting solenoidal coil into the design of Ref. \cite{13}.

The experience accumulated in two decades with ICARUS is very important.
 However, a substantial R\&D is required, at an extent which depends on the
 detector design and on the (underground) location.
 In the design of very large detectors, with or without magnetic field,
 one has to proceed from concepts or conceptual designs to a practical design.

\subsection{The Emulsion Cloud Chamber}

The use of Emulsion Cloud Chamber (ECC) has been proposed to detect "silver events"
from $\nu_e - \nu_{\tau}$ oscillations at a $\nu$-Factory, in order to
complement the $\nu_e - \nu_{\mu}$ "golden events" in resolving
$\theta_{13} - \delta$ ambiguities \cite{18}.

The ECC of the OPERA experiment
\cite{OPERA} in the CNGS beam consists of a multiple sandwich of lead and nuclear
emulsion sheets. The production and decay
of $\tau$ leptons is expected to be observed with a very low background thanks to the emulsion 
sub-$\mu$m resolution. The 1.8 kton OPERA target is built up with about 200,000
lead-emulsion "bricks". 

At the $\nu$-Factory, a 4 kton mass is envisaged. By scanning only
"wrong sign" muons as those coming from $\tau$ decays, the scanning load is expected 
to be comparable with that in OPERA. An increase in the speed of the automatic microscopes is foreseeable. The OPERA Collaboration is working to provide a milestone in the application of the ECC technique.

\section{Preliminary comparison of accelerator facilities }

The comparison of performances of different facilities cannot be considered as concluded. 
Several different aspects still need to be clarified before a final comparison can be performed.
\begin{itemize}
	\item
Costs, timescales, fluxes of the different accelerators systems are not yet fully 
worked out.
	\item
Performances and optimization of the detectors are not known 
at the same level: 
for water \v{C}erenkov detectors full simulation and full reconstructions 
of the events are available, 
based on the experience of Super-Kamiokande, but the optimization  
in terms of photo-detector coverage is still to be performed. 
The optimization of the magnetic detector (and of the assumed cuts) 
for the Neutrino Factory was performed for very small values of $\theta_{13}$, 
and the performances are for the moment 
based on parametrization of the MINOS performances,
and very low values of backgrounds from hadron decays (pions, kaons, and charm)
are claimed possible.   
The performances of the emulsion detector for the silver channel of Nufact 
or of the liquid argon detector are based on full simulations, 
but will need to be bench-marked using the performances of OPERA and ICARUS 
respectively.
	\item
Several different measurements can be defined as significant for the facility, 
and they cannot be optimized all together (see also reference \cite{HuberBB}).
For instance the following measurements bring to different optimizations:
sensitivity to \thetaot, discovery of subleading \numunue\  oscillations,
unambiguous measurement of \thetaot,
 measure of
\sigdm, discovery of leptonic CP violation, unambiguous measurement of  \delCP.
	\item
The final extraction of the oscillation parameters can significantly change depending on technical aspects of the fitting programs (like choice of the input parameters, treatment of the errors on the neutrino oscillation parameters, treatment of degeneracies etc.).

 The GLoBES program \cite{Globes} represents a major improvement:
 it allows to compare different facilities keeping the same the fitting program,
 and it makes explicit the description of the performances of the detectors.
 We strongly recommend that new developments in this field will make use of
GLoBES, in view of a more transparent comparison of the different proposals.
	\item
Systematic errors that strongly influence performances, for instance sensitivity to
leptonic CP violation for large values of \thetaot, are not substantially discussed in the literature. We are confident that facilities where neutrino fluxes can be known a-priori,
as the case of Beta Beams and Neutrino Factories, will have smaller systematic errors
(and smaller backgrounds) than e.g. neutrino SuperBeams. 
This difference is
not known quantitatively today.

The concept of the near detector station(s) and flux monitoring 
systems has to be proposed together
with the facility, in particular for low energy (few 100 MeV) 
BetaBeam and SuperBeam where
the issues of muon mass effect, Fermi motion 
and binding energy uncertainty are 
highly non-trivial.

Finally, for the Neutrino Factory, 
the question of systematics on the prediction 
of matter effects is essential for the performance at large values of \thetaot.
	\item
Overall performances will depend on the combination of several different inputs.
For instance low energy Super Beams and Beta Beams can profit of atmospheric neutrino
oscillations, detected with a large statistics in the gigantic water \v{C}erenkov detector,
to solve degeneracies and measure $sign(\Delta m^2_{23})$, as shown in the  \cite{Schwetz}.
Neutrino Factory can profit of the combination of different channels as the 
golden and the silver one (see Section \ref{sec:Nufact}), detectors at different baselines,
atmospheric neutrinos collected in the the iron magnetized detector \cite{Indiani}.
A full exploration of these possibilities is an ongoing process and the results available at today cannot be considered final.
\end{itemize}

\begin{figure}
    \epsfig{file=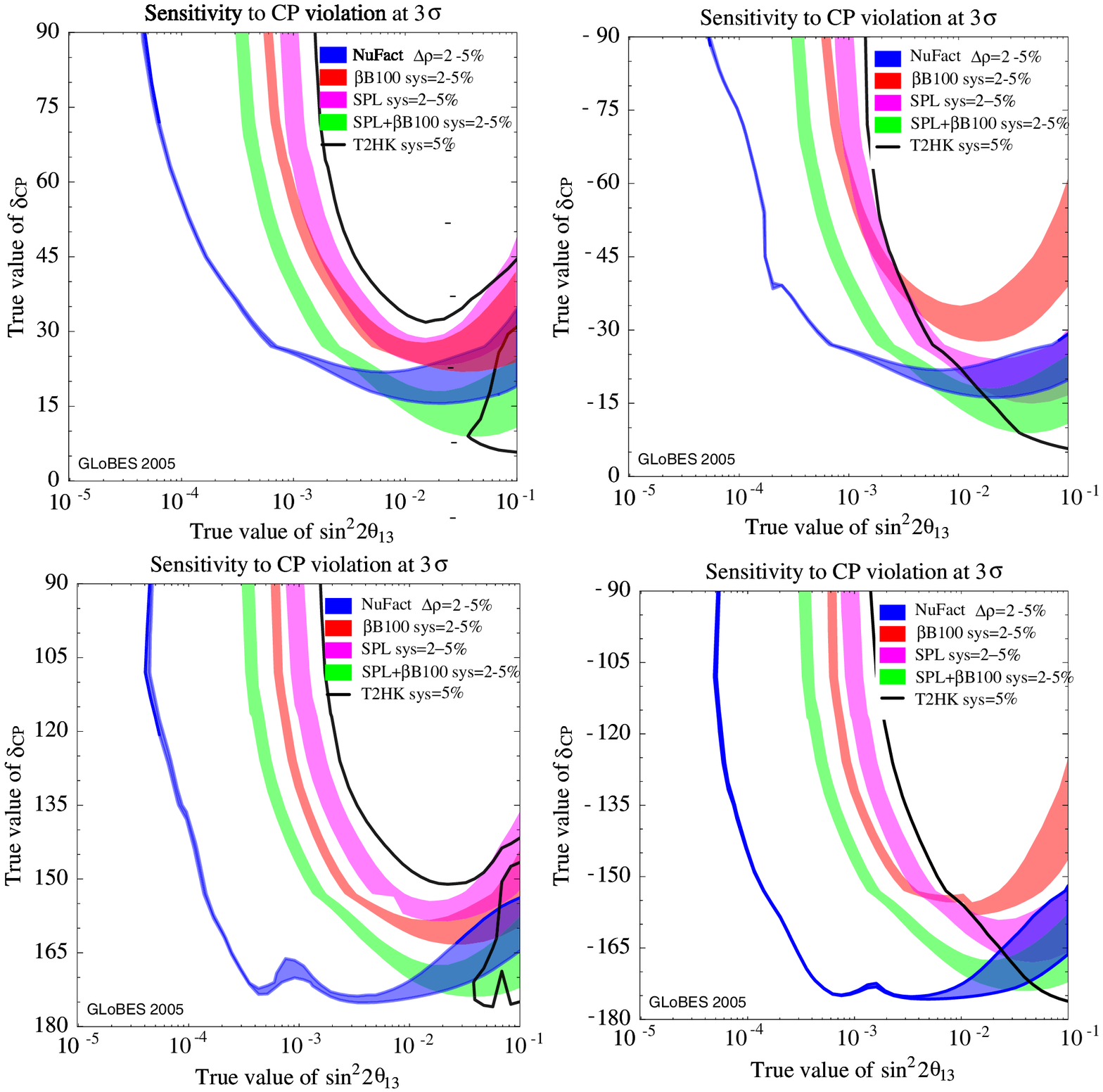,width=\textwidth}  
    \mycaption{
      $\delCP$ discovery potential at $3 \sigma$ (see text) computed
      for 10 years running time. For explanation of the facilities see the text.     The four plots represent the four possible quadrants of \delCP values,
   performances of the different facilities are not at all the same in the different quadrants.
 The width of the curves reflects the range of systematic errors: 2\% and 5\% on signal and background errors for SPL-SB and Beta Beam, 2\% and 5\% for the matter density. Other systematic errors are 5\% on signal and background of T2HK,
 0.1\% for nufact signal, 20\% for nufact backgrounds. Description of the facilities can be found in 
      \cite{Globeslib} }
  \label{fig:final}
\end{figure}

Having said that, a comparison of the facilities that at present are described in
the GLoBES library \cite{Globeslib}, as far as concerns 
and leptonic CP violation discovery potential, is shown in Fig.~\ref{fig:final}.

The plots show the sensitivity to CP violation at 3$\sigma$ CL
($\Delta\chi^2=9$). Sensitivity to CP violation is defined, for a given
point in the \thetaot-\delCP\ plane, by being able to exclude \delCP=0 and
$\delCP=\pi$ at the given confidence level. All plots have been prepared
with GLoBES \cite{Globes}.

Degeneracies and correlations are fully taken into account. For all
setups the appropriate disappearance channels have been included. The
Beta Beam is lacking muon neutrino disappearance, but the result does
not change if T2K disappearance information is included in the
analysis. In all cases systematics between neutrinos, anti-neutrinos,
appearance and disappearance is uncorrelated. For all setups with a
water \v{C}erenkov detector the systematics applies both to background
and signal, uncorrelated.

The Neutrino Factory assumes $3.1 \cdot 10^{20} \mu^+$ decays per year for 10
years and $3.1 \cdot 10^{20} \mu^-$ decays for 10 years.
 It has one detector with
m=100 kton at 3000 km and another 30 kton detector at 7000 km. The
density errors between the two baselines are uncorrelated, sensitivities
are computed for 2\% and 5\% systematic error on matter density. The
systematics are 0.1\% on the signal and 20\% on the background,
uncorrelated. The detector threshold and the other parameters are
taken from \cite{Huber02} and approximate the results of 
\cite{Cervera:2000vy}.

The Beta Beam assumes $5.8 \cdot 10^{18}$ \He decays per year for five years and
$2.2 \cdot 10^{18}$ \Ne decays per year for five years. The detector mass is
500 kton. The detector description and the globes-file is from 
\cite{MMNufact05}.

The SPL setup is taken from \cite{Campagne}, and the detector mass is 500 kton.

The T2HK setup is taken from \cite{Huber02}
  and closely follows the LoI \cite{T2K}. The
detector mass is 1000 kton and it runs with 4MW beam power, 6 years with
anti-neutrinos and 2 years with neutrinos. The systematic error on
both background and signal is 5\%.

The oscillation parameters were \cite{Valle,Lisi2}:
$\delta m^2_{23}=0.0024\, eV^2$,
$\delta m^2_{12}=0.00079\, eV^2$,
$\thetatt=\pi/4$,
$\theta_{12}=0.578$.
The input errors are (at 1 $\sigma$):
10\% on $\delta m^2_{23}$,
10\% on \thetatt,
10\% on $\theta_{12}$,
 4\% on $\delta m^2_{12}$,
 5\% on $\rho$ (unless otherwise stated).


\end{document}